\documentclass[10pt,letterpaper]{article}
\usepackage[top=0.85in,left=2.75in,footskip=0.75in]{geometry}

\usepackage{amsmath,amssymb}

\usepackage{changepage}

\usepackage[utf8x]{inputenc}

\usepackage{textcomp,marvosym}

\usepackage{cite}

\usepackage{nameref,hyperref}

\usepackage[right]{lineno}

\usepackage{microtype}
\DisableLigatures[f]{encoding = *, family = * }

\usepackage[table]{xcolor}

\usepackage{array}

\newcolumntype{+}{!{\vrule width 2pt}}

\newlength\savedwidth

\usepackage{setspace} 

\raggedright
\usepackage[aboveskip=1pt,labelfont=bf,labelsep=period,justification=raggedright,singlelinecheck=off]{caption}

\makeatletter
\renewcommand{\@biblabel}[1]{\quad#1.}
\makeatother

\usepackage{lastpage,fancyhdr,graphicx}
\usepackage{epstopdf}
\fancyheadoffset[L]{2.25in}
\fancyfootoffset[L]{2.25in}
\usepackage{graphicx} 
\usepackage{soul} 
\usepackage{booktabs}
\usepackage{tabularx}
\usepackage{colortbl}
\usepackage{url}

\usepackage[strings]{underscore}

\usepackage{hyperref}
\usepackage{amsfonts}
\usepackage{nicefrac}
\usepackage{microtype}
\usepackage{lipsum}
\usepackage{tabularx}
\usepackage{caption} 
\usepackage{float}
\usepackage{subcaption}
\usepackage{caption}
\usepackage{dcolumn}
\usepackage{booktabs}
\usepackage{adjustbox}
\usepackage{bbm}
\usepackage{amsmath}

\usepackage{multirow}

\usepackage{marvosym}

\usepackage{color}
\definecolor{lightgray}{gray}{0.90}
\usepackage{soul}

\usepackage{etoolbox}
\newtoggle{NOFIG}
\togglefalse{NOFIG} 

\newcommand\greybox[1]{%
  \vskip\baselineskip%
  \par\noindent\colorbox{lightgray}{%
    \begin{minipage}{\textwidth}#1\end{minipage}%
  }%
  \vskip\baselineskip%
}

\begin{document}
\vspace*{0.2in}

\begin{flushleft}
{\Large
\textbf\newline{Structural invariants and semantic fingerprints in the ``ego network" of words} 
}
\newline
\\
Kilian Ollivier\textsuperscript{1*},
Chiara Boldrini\textsuperscript{1},
Andrea Passarella\textsuperscript{1},
Marco Conti\textsuperscript{1} 
\\
\bigskip
\textbf{1} CNR-IIT, Pisa, Italy
\\
\bigskip

%
%





* Corresponding author

E-mail: kilian.ollivier@iit.cnr.it (KO)

\end{flushleft}

%
%
%
%
%

\section*{Abstract}

Well-established cognitive models coming from anthropology have shown that, due to the cognitive constraints that limit our ``bandwidth'' for social interactions, humans organize their social relations according to a regular structure.
In this work, we postulate that similar regularities can be found in other cognitive processes, such as those involving language production. In order to investigate this claim, we analyse a dataset containing tweets of a heterogeneous group of Twitter users (regular users and professional writers).
Leveraging a methodology similar to the one used to uncover the well-established social cognitive constraints, we find regularities at both the structural and semantic levels. In the former, we find that a concentric layered structure (which we call \emph{ego network of words}, in analogy to the ego network of social relationships) very well captures how individuals organise the words they use. The size of the layers in this structure regularly grows (approximately 2-3 times with respect to the previous one) when moving outwards, and the two penultimate external layers consistently account for approximately 60\% and 30\% of the used words, irrespective of the number of layers of the user. For the semantic analysis, each ring of each ego network is described by a semantic profile, which captures the topics associated with the words in the ring. We find that ring \#1 has a special role in the model. It is semantically the most dissimilar and the most diverse among the rings. We also show that the topics that are important in the innermost ring also have the characteristic of being predominant in each of the other rings, as well as in the entire ego network. In this respect, ring \#1 can be seen as the semantic fingerprint of the ego network of words. 


\section{Introduction}
\label{sec:intro}

In humans, language production is a deliberate and conscious action. However, it relies on many invisible mental processes that allow the construction of sentences in a very short time. For example, these cognitive processes are at play during the word retrieval stage, when the brain has to efficiently process, in a few milliseconds, its lexicon in order to find the right word, among thousands of others, that best fits the concept that needs to be expressed~\cite{levelt1999theory}. In order to achieve this impressive performance, cognitive strategies that exploit language properties, such as word frequency (e.g. when the most frequently used words are retrieved more quickly~\cite{broadbent1967word,qu2016tracking}), are activated. In this paper, we set out to find traces of these cognitive patterns in written production with a data-driven approach. To this end, we rely on the ego network model, which has already uncovered the cognitive limits of another human activity: socialisation. 

\subsection{The social ego network model}

\begin{figure}[h]
\begin{center}
\iftoggle{NOFIG}{}{\includegraphics[scale=0.7]{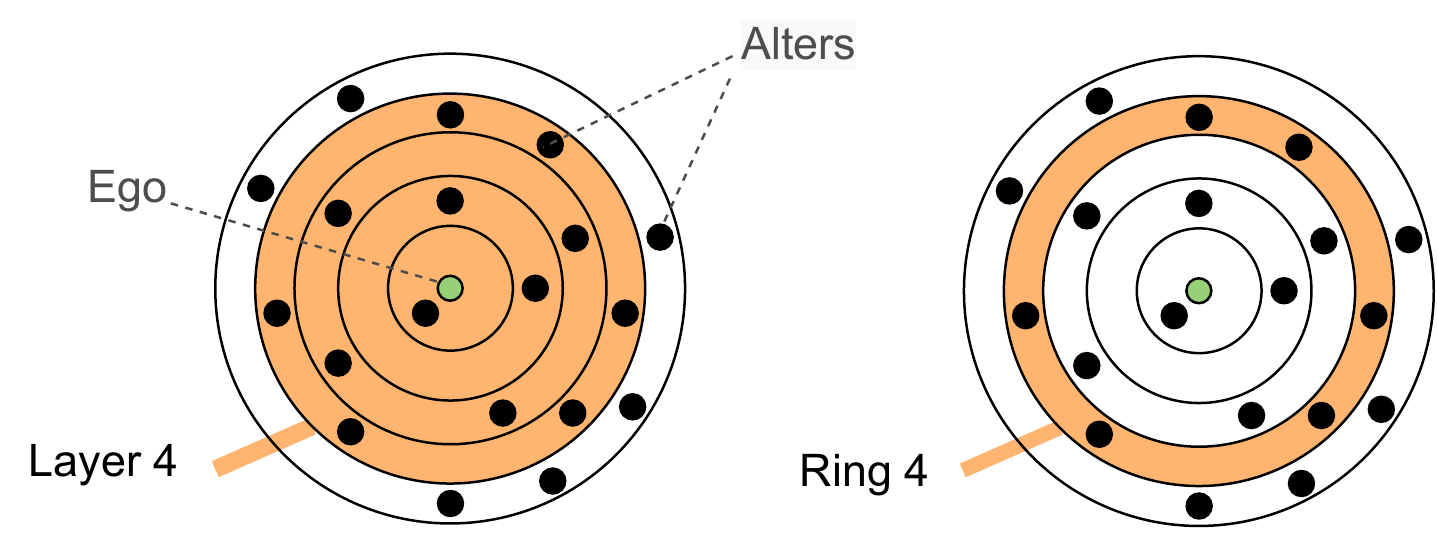}}
\caption{{\bf The \emph{ego network} of social relationships.} The green dot symbolizes the ego and the black dots the alters with whom the ego maintains an active social relationship. A layer also contains the alters of the inner layers, unlike the rings.
}
\label{fig:egonet}
\end{center}
\end{figure}

Anthropologists have shown that the number of meaningful social relationships that humans can maintain is not only limited to 150~\cite{dunbar1998social} (the famous Dunbar's number) but it is also stable over time. The discovery of this regularity in human activity stems from the observation that, in different species of primates, there exists a correlation between the size of the neocortex (the part of the brain dedicated to high-level cognitive functions such as socialisation, language, etc.), and the average size of groups in natural environments. Extrapolating the expected size of a human group from the dimension of the human brain, as well as studying historical data such as the maximum size before fission of autonomous communities~\cite{Dunbar.2018}, the Dunbar number consistently emerges. It was then shown that these 150 active social relationships can be further subdivided into 4 \emph{concentric} circles~\cite{hill2003social,Zhou2005}, the innermost one containing the most intimate social relationships~\cite{dunbar2015structure}, the outermost one enclosing all 150 social relationships. The typical size of these concentric circles is 5, 15, 50, and 150, respectively, with a constant scaling ratio of about 3 between consecutive circles. Note that the portion of a circle not included in its innermost ones is referred to as \emph{ring}. This hierarchical structure of social relationships is called ``ego network''. Recent studies based on data collected from online social networks have shown that online relationships are subject to the same laws as offline ones: the size of the ego network (i.e., the total number of social relationships) remains in the same order of magnitude as the Dunbar's number, which indicates that the cognitive constraint yielding this number is not overridden by a communication medium that facilitates social interactions~\cite{dunbar2015structure,Haerter2012,Miritello2013,gonccalves2011modeling}. In OSNs (Online Social Networks), the typical number of circles is slightly higher than 4, due to the presence of an additional circle in the center of the ego network (containing about 1.5 people), but the scaling ratio is preserved at around 3 (Fig.~\ref{fig:egonet}). 

\subsection{From social ego networks to ego networks of words}

The ego network model highlights the regularity of the structure of social relations, in real life and in OSN. In this paper, we adopt an analogous approach to investigate the regularities and invariants manifesting cognitive constraints in language production. Specifically, we conjecture that a similar structure, which we call \emph{``ego network of words"}, may also be used to describe the way humans use words, and that this structure may provide very significant information to characterise the peculiarities of individuals, similarly to the social dimension. In fact, it is known~\cite{sutcliffe2012relationships} that many traits of social behavior (resource sharing, collaboration, diffusion of information) are chiefly determined by the structural properties of social ego networks.

The motivation for this analogy is twofold. First, the use of words is, much like socialisation, a process that involves the use of cognitive resources, thus we conjecture that the ego network model may have larger applicability in describing how humans allocate cognitive resources, for example to language. Second, language is a social activity, whose emergence is potentially linked to the surge in active human relationships from the 50 of the closest primate to 150 for humans. This theory, known as \emph{social gossip theory of language evolution}~\cite{Dunbar1998}, postulates that language facilitates grooming social relations by reaching several peers at the same time. In addition, there is already well-established knowledge of a number of empirical cognitive limits affecting language, such as the bounded size of our vocabulary (which is consistently limited to approximately $42,000$ words for a native 20-year-old English speaker~\cite{Brysbaert2016}), as well as the Zipf's law of words~\cite{zipf1949human}, which states that the frequency of a word is inversely proportional to its position in the frequency table for most human writings. We, therefore, choose to study the individual distribution of vocabulary, by forming concentric circles of words according to their frequency of use by the ego in question. Then, going beyond words as units of language, we focus on the topics to which the words refer.  We thus complement the structural analysis with a semantic study, which completes our cognitive analysis framework. In the same way that the social ego network model has been used to provide a different perspective to social network analysis (such as for information diffusion~\cite{arnaboldi2017online}), we want to leverage the ego networks of words as microscopes to discover novel properties of language production.

\subsection{Contribution and key findings}

The main contribution of this work is the structural and semantic analysis of the ego networks of words for Twitter users. 
By using the ego network model, in this paper, we uncover complex structures showing that the cognitive effort to organise one's vocabulary is limited in many ways. We choose a corpus of text made up of tweets because it allows us to work with a varied sample of ``authors'' (e.g. more varied than a corpus of newspaper articles). Moreover, as Twitter is dedicated to the exchange of very short messages (240 characters), it is a medium that is very favourable to spontaneous reactions, with a more natural style and a reduced writing time. This time constraint is more likely to reveal human behaviour, in analogy with the social domain, where time limitations have been shown to significantly affect social cognitive constraints~\cite{Dunbar1998}. 
%
For our data-driven analysis, we collected tweets from generic as well as specialised Twitter users (Section~\ref{sec:dataset}). Using the ego-network-of-words model, we are able to find evidence of a structural regularity in the frequency of word usage by each individual (Section~\ref{sec:structural}). The semantic analysis (Section~\ref{sec:semantic}) also establishes the existence of additional invariants, but most importantly it uncovers the nature of the innermost layer as the \emph{semantic fingerprint} of the whole ego network, i.e., this layer groups together the most important topics on which the user is active. This strengthens the analogy with the social version of the ego network model, where the innermost layers include the most important social relationships of a person.

\noindent
The key findings of the paper are the following.
\begin{itemize}
    \item Similarly to the social case, we found that a \emph{regular concentric, layered structure} (which we call \emph{ego network of words} in analogy to the ego networks of the social domain) very well captures how an individual organizes their cognitive effort in language production. Specifically, words can be typically grouped in between 5 and 7 layers of decreasing usage frequency moving outwards, regardless of the specific class of users (regular vs professional).
    \item One structural invariant is observed for the \emph{size of the layers}, which approximately doubles when moving from layer $i$ to layer $i+1$. The only exception is the innermost layer, which tends to be approximately 5 five times smaller than the next one. This suggests that the innermost layer, the one containing the most used words, may be drastically different from the others.
    \item A second structural invariant emerges for the \emph{external layers}. Users with more layers organise differently their innermost layers, without modifying significantly the size of the most external ones. In fact, while the size of all layers beyond the first one linearly increases with the most external layer size, the second-last and third-last layers consistently account for approximately 60\% and 30\% of the used words, irrespective of the number of layers of the user.
    \item The semantic analysis of the words contained in the ego networks confirms that layer~\#1 is exceptional in the ego networks of words: it generates proportionally more topics than the other rings,
    these topics are more diverse, and its overall semantic profile is the most different with respect to those of other rings.
    \item In addition, topics that are important in ring \#1 tend to be important in other rings as well (we call this the \emph{pulling power} of ring \#1). Thus, layer~\#1, despite being the smallest, can be seen as the \emph{semantic fingerprint} of the ego network of words.
    \item The topics that are primary in some rings tend to be stronger than average among the primary and non-primary topics in the semantic profile of the other rings. This shows that, while layer \#1 provides a particularly strong signal about prevalence in the ego networks, weaker signals show a more complex structure of influence among topics ``resident" in different layers of the ego network of words.
\end{itemize}

This paper extends our prior publication in~\cite{ollivier2020}, where the structural analysis was carried out. Specifically, in this paper, we also present an extensive semantic analysis of the ego network of words. This allows us to provide a much more comprehensive understanding of the model, and highlight ways to characterise specificities of individuals as they emerge from their use of words, in addition to structural invariants observed through the structural properties of the ego networks.

\section{Related work}

To the best of our knowledge, no work has been published yet on models of individual word organisation similar in spirit to ours (i.e., by exploring the analogy with the social ego network model). However, some work has already been done on individual word frequency distribution by extending the notion of Zipf’s law~\cite{piantadosi2014zipf}. Based on Zipf’s law, some have tried to find a generative model that could explain such a regularity-based human cognition~\cite{anderson1991reflections}, or just how the limited capacities of our memory naturally constrain our long-term use of words~\cite{graesser1978limited}. More generally, vocabulary size is often studied in the context of language learning for both children and adults, as well as to detect possible cognitive impairments~\cite{aramaki2016vocabulary}. For the semantic part, we have not identified any previous work on modelling user interests with a stratified approach, such as ours, that relies on the ego network of words. Most publications are about topic recommendations (relying upon a wide range of techniques, such as hashtag analysis~\cite{abel2011analyzing}, LDA~\cite{bhattacharya2014inferring} or ontology databases~\cite{frasincar2009semantic}), and about the emergence and monitoring of trending topics on Twitter~\cite{arslan2022understanding,guille2014mention}.

\section{The dataset}
\label{sec:dataset}

The analysis is built upon four datasets extracted from Twitter, using the official Search and Streaming APIs (note that the number of downloadable tweets -- at the time of download -- was limited to the most recent 3200 tweets per user). Each of them is based on the tweets issued by users in four distinct groups:
\begin{description}
\item[Journalists] Extracted from a Twitter list containing New York Times journalists (\url{https://twitter.com/i/lists/54340435}), created by the New York Times itself. It includes 678 accounts, whose timelines have been downloaded on February 16th, 2018. 
\item[Science writers] Extracted from a Twitter list created by Jennifer Frazer (\url{https://twitter.com/i/lists/52528869}), a science writer at \textit{Scientific American}. The group is composed of 497 accounts and has been downloaded  on June 20th, 2018.
\item[Random users \#1] This group has been collected by sampling among the accounts that posted a tweet or a retweet in English with the hashtag \textit{\#MondayMotivation} (at the download time, on January 16th, 2020). This hashtag is chosen in order to obtain a diversified sample of users: it is broadly used and does not refer to a specific event or a political issue. This group contains 5183 accounts after bot filtering.
\item[Random users \#2] This group has been collected by sampling among the accounts  that posted a tweet or a retweet in English, from the United Kingdom (we set up a filter based on the language and country), at download time on February 11th, 2020. This group contains 2733 accounts after bot removal. 
\end{description} \vspace{-5pt}

These four groups are chosen to cover different types of users: the first two contain accounts that use language professionally (journalists and science writers) and the other two contain regular users, which are expected to be more colloquial and less controlled in the language they use.
Since the random user accounts are not handpicked as in the two first groups, we need to make sure that they represent real humans. The probability that an account is a bot is calculated with the Botometer service~\cite{davis2016botornot}, which implements a state-of-the-art bot detection algorithm. This probability that the account is not human, which is called ``complete automation probability'' (CAP), is  not only based on linguistic features such as grammatical tags, or the number of words in a tweet, but also on language-agnostic features like the number of followers or the tweeting frequency~\cite{varol2018feature}. There is no standard CAP threshold to easily separate bots from humans: it depends on the expected balance of precision and recall. That is why we discard accounts with a CAP higher than 0.5, which considerably limits the number of false negatives (undetected bots). The Botometer service achieves a performance of 0.95 AUC on standard bot detection datasets~\cite{davis2016botornot}. With this configuration, the algorithm detects 29\% of bot accounts in the dataset of random users\#1 and 23\% in the dataset of random users\#2. 

In our analysis, we only consider the timelines of \emph{active} Twitter accounts, i.e., users that tweet regularly. Since this preprocessing step largely follows the standard approach in the related literature~\cite{Dunbar2015,boldrini2018twitter}, further details are left to the~\nameref{S1-Appendix}. 
Please note that we discard retweets with no associated comments, as they do not include any text written by the target user, and tweets written in a language other than English (since most of the NLP tools needed for our analysis are optimised for the English language).
%


\subsection{Extracting user timelines with the same observation period}
\label{sec:observation-period}

As discussed above, for each user in our datasets we retrieved the most recent 3200 tweets (due to the Twitter API limitation), which constitute the \emph{observed timeline} of the user. The time period covered by these tweets varies according to the frequency with which the account is tweeting: for very active users, the last 3200 tweets will only cover a short time span. Since random users are generally more active, their observation period is shorter
, and this may create a significant sampling bias. In fact, the length of the observation period affects the measured word usage frequencies (specifically, we cannot observe frequencies lower than the inverse of the observation period). In order to guarantee a fair comparison across user categories and to be able to compare users with different tweeting activities without introducing biases, we choose to work on timelines with the same duration, by restricting to an observation window $T$. To obtain timelines that have the same observation window $T$ (in years), we delete all those with a duration shorter than $T$ and remove tweets written more than $T$ years ago from the remaining ones. 

\begin{figure}[h]
\centering
  \centering
  \iftoggle{NOFIG}{}{\includegraphics[width=0.6\linewidth]{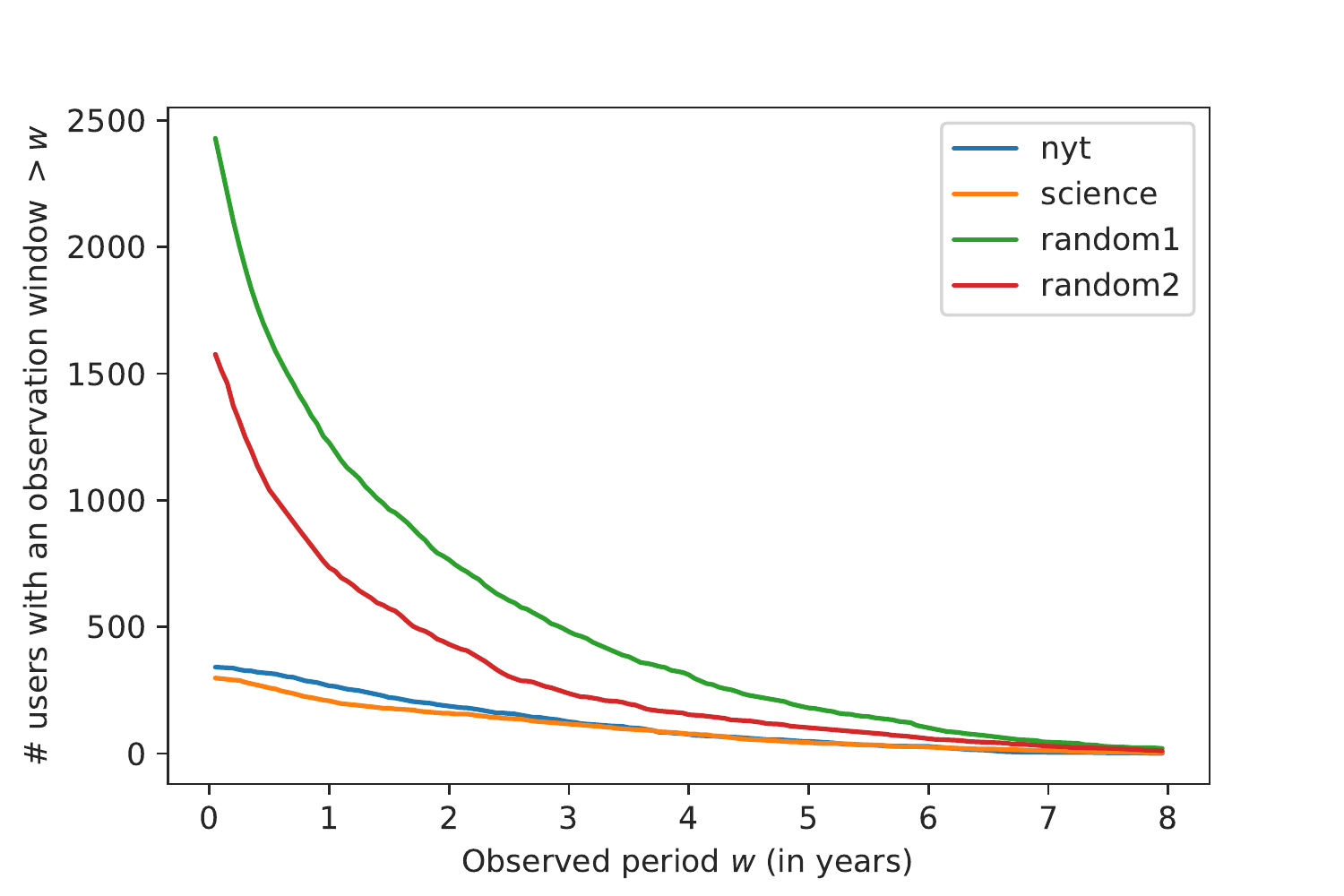}}
  \captionof{figure}{{\bf Available timelines.} Number of selected timelines depending on the observation window.}
  \label{fig:nb-users}
 \end{figure}

\begin{figure}[h]
  \centering
  \iftoggle{NOFIG}{}{\includegraphics[width=0.6\linewidth]{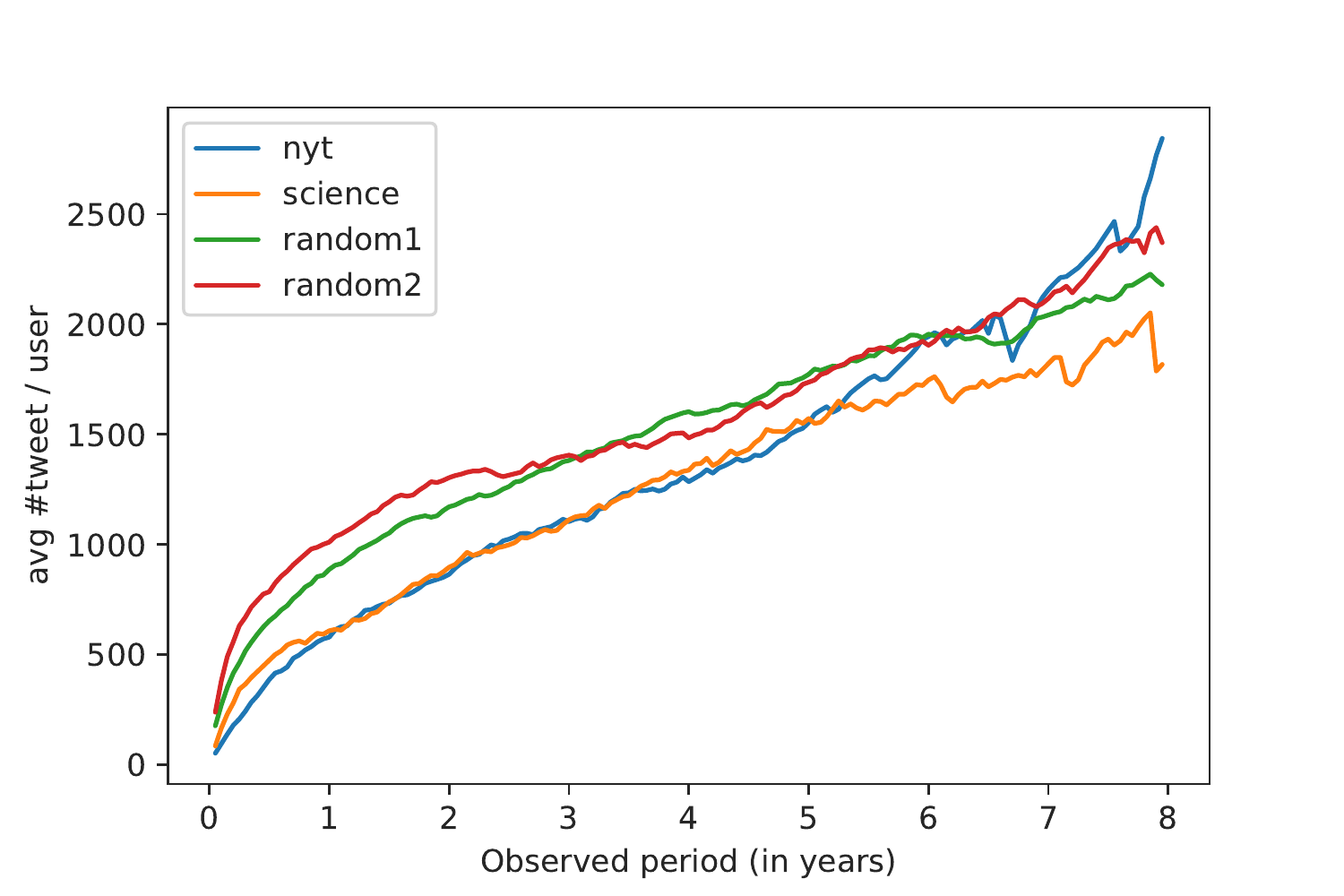}}
  \captionof{figure}{{\bf Tweets per user.} Average number of tweets depending on the observation window. The Pearson linear correlation coefficient is equal to or greater than $.98$ for the four datasets.}
  \label{fig:avg-tweet}
\end{figure}

Increasing $T$ reduces the number of users we can keep for our analysis (see Fig.~\ref{fig:nb-users}): for a $T$ larger than 2 years, that number is halved, and for a $T$ larger than 3 years, it falls below 500 for all datasets. On the contrary, the average number of tweets per timeline increases linearly with $T$ (Fig.~\ref{fig:avg-tweet}). The choice of an observation window will then result from a trade-off between a high number of timelines per dataset and a large average number of tweets per timeline. To simplify the choice of $T$, we only select round numbers of years. We can read in Table~\ref{tab:volume-same-period} that, beyond 3 years, the number of users falls below 100 for some datasets. On the other hand, the number of tweets for $T = 1 \textrm{ year}$ remains acceptable ($> 500$). Since we value the diversity of users (in order to limit any bias in the selection of Twitter accounts) over the number of tweets available, we make the choice of $T = 1 \textrm{ year}$ for the entire paper.  Results with other $T$ lengths can be found in~\cite{ollivier2020}. We note that random users have a higher frequency of tweeting than others. This difference tends to smooth out when the observation period is longer (Table~\ref{tab:volume-same-period}). This can be explained by the fact that the timelines with the highest tweeting frequency are excluded in that case because their observation period is too small (which further supports the fact that a smaller $T$ reduces the selection bias of users).

%

\begin{table}[!ht]
\centering
\caption{{\bf Datasets summary}. Number of users and tweeting frequency at different observation windows. \label{tab:volume-same-period}}
\scriptsize
\setlength{\tabcolsep}{0.5em}
\renewcommand{\arraystretch}{1.2}
    \begin{tabular}{lcccccc}
        \toprule
        \multirow{2}{*}{Datasets} & \multicolumn{3}{c}{Number of users} & \multicolumn{3}{c}{Avg \# of tweets / user} \\ \cline{2-7} 
        & 1 year & 2 years & 3 years & 1 year & 2 years & 3 years \\ \midrule
            NYT Journalists & 268 & 187  & 125 & 579.71 & 865.02 & 1104.58 \\
            Science Writers & 208 & 159 & 117 & 609.08 & 897.29 & 1112.63\\
            Random Users \#1 & 1227 & 765 & 311 & 897.29 & 1179.98 & 1403.50\\
            Random Users \#2 & 734 & 431 & 153 & 1057.41 & 1315.71 & 1404.60\\
        \bottomrule
    \end{tabular} 
\end{table}
 
\section{Structural analysis of the ego network of words}
\label{sec:structural}

In this section, we focus on the analysis of structural properties of the ego network of words, highlighting structural invariants in language production. Note that, in the social domain, pure structural properties of ego networks were instrumental~\cite{sutcliffe2012relationships} in characterising many traits of social behavior (resource sharing, collaboration, diffusion of information). For this reason, we believe it is important to assess them in the language domain as well, before moving on (Section~\ref{sec:semantic}) to more complex and domain-specific analyses.

We first describe the methodology we use for our analysis in Section~\ref{sec:structural-methods}, then we discuss the results in Section~\ref{sec:structural-results}. For ease of reading, the notation used in this section is summarised in Table~\ref{tab:notation-structural}. The section reports only the most significant results obtained by analysing the structural properties of the ego network. Interested readers are referred to~\cite{ollivier2020} for additional results.

\begin{table}[!ht]
\centering
\caption{{\bf Summary of notation used in the structural analysis}}
\footnotesize
\begin{tabular}{@{}lcp{6cm}@{}}
\toprule
\textbf{Name} & \textbf{Notation} & \textbf{Definition/formula}\\
\midrule
Optimal number of circles &~$\tau^{(e)}$ & the results of the clustering on the word frequencies for the user (ego)~$e$ \\
Circle (or layer) &~$\mathcal{L}_i^{(e)}$ &~$i$-th social circles of the tagged ego~$e$, with~$i \in \{1, \ldots, \tau^{(e)}\}$ \\
Scaling ratio of layer~$i$ &~$\rho_i^{(e)}$ &~$\frac{|\mathcal{L}_{i}^{(e)}|}{|\mathcal{L}_{i-1}^{(e)}|}$, with~$i \in \{2, \ldots, \tau^{(e)}\}$ \\
Ring  &~$r_i^{(e)}$ &~$\mathcal{L}_i^{(e)} - \mathcal{L}_{i-1}^{(e)}$ \\
\bottomrule
\end{tabular}
\label{tab:notation-structural}
\end{table}

\subsection{Methods}
\label{sec:structural-methods}

For each user, acting as ego, we want to build their ego network of words. To this aim, we first extract individual words from the user's tweets (Section~\ref{sec:structural-methods-words}), then we build the actual ego network from these words (Section~\ref{sec:structural-methods-egonet}). 

\subsubsection{Word extraction}
\label{sec:structural-methods-words}

Since the analysis focus on words and their frequency of use, we take advantage of NLP techniques for extracting them. As a first step, all the syntactic marks that are specific to communication in online social networks (mentions with~@, hashtags with \#, links, emojis) are discarded (see
\nameref{S1-Appendix}
for a summary). 
Once the remaining words are tokenized (i.e., identified as words), those that are used to articulate the sentence (e.g., ``with", ``a", ``but") are dropped. In linguistics, this type of word is called a functional word as opposed to lexical words, which have a meaning independent of the context. These two categories involve different cognitive processes (syntactic for functional words and semantic for lexical words), different parts of the brain~\cite{diaz2009comparison}, and probably different neurological organizations~\cite{friederici2000segregating}. We are more interested in lexical words because their frequency in written production depends on the author's intentions, as opposed to functional word frequencies that depend on language characteristics. Functional words may also depend on the style of an author (and due to this they are often used in stylometry). Still, whether their usage requires a significant cognitive effort is arguable, hence in this work, we opted for their removal. Moreover, lexical words represent the biggest part of the vocabulary. Functional words are generally called stop-words in the NLP domain and we simply used an already existing list from the library spaCy~\cite{spacy2} to remove them.

As this work will leverage word frequencies as a proxy for discovering cognitive properties, we need to group words derived from the same root (e.g. ``work" and ``worked") in order to calculate their number of occurrences. This operation can be achieved with two methods: stemming and lemmatization. Stemming algorithms generally remove the last letters thanks to complex heuristics, whereas lemmatization uses the dictionary and a real morphological analysis of the word to find its normalized form. Stemming is faster, but it may cause some mistakes in overstemming and understemming. For this reason, we choose to perform lemmatization with the help of the package WordNetLemmatizer from the library NLTK~\cite{loper2002nltk} (which leverages the lexical database WordNet). 
Once we have obtained the number of occurrences for each word base, we remove all those that appear only once to leave out the majority of misspelled words.
The~\nameref{S1-Appendix}
contains examples of the entire preprocessing part. 

In the remaining of the paper, when we talk about the ``words'' of a user, we refer to the set of words left after removing functional words and after lemmatization. 

\subsubsection{Building the ego network of words}
\label{sec:structural-methods-egonet}

Let us focus on a user~$j$. When studying the social cognitive constraints~\cite{Dunbar2015}, the contact frequency between two people was taken as a proxy for their intimacy and, as a result, for their cognitive effort in nurturing the relationship. Similarly, the frequency~$f_i$ at which user~$j$ uses word~$i$ is considered here as a proxy of their ``relationship". Frequency~$f_i$ is given by~$\frac{n_{ij}}{T}$, where~$n_{ij}$ denotes the number of occurrences of word~$i$ in user~$j$'s timeline, and~$T$ denotes the observation window of~$j$'s timeline in years ($T=1y$ in our case, as discussed in Section~\ref{sec:observation-period}). 
Using this frequency definition, we now investigate whether the words of a user can be grouped into homogeneous classes and whether different users feature a similar number and sizes of classes. To this aim, for each user, we leverage a clustering algorithm to group words with a similar frequency. 
The selected algorithm is Mean Shift~\cite{fukunaga1975estimation}, because as opposed to Jenks~\cite{jenks1977optimal} or k-means~\cite{macqueen1967some}, it is able to automatically detect the optimal number of clusters. 
In order to account for the long-tailed nature of frequencies, a standard log-transformation is applied to the frequency values prior to the Mean Shift run.

Thus, for each user, we feed the user's words to Mean Shift. The output of the clustering process is one value~$\tau^{(e)}$ for each ego network~$e$, which describes the optimal number of classes (clusters) in which the word frequencies can be split. We rank each cluster by its position in the frequency distribution: cluster \#1 is the one that contains the most frequent words, and the last cluster is the one that contains the least used words. Following the convention of the social ego network model discussed in Section~\ref{sec:intro}, these clusters can be mapped into concentric layers (or circles), which provide a cumulative view of word usage. Specifically, layer~$\mathcal{L}_i$ includes all clusters from the first to the~$i$-th. Layers provide a convenient grouping of words used \emph{at least} at a certain frequency. We refer to this layered structure as the \emph{ego network of words}.  Note that, since layers in ego networks are cumulative (i.e., they include all words used at least a certain frequency), we will use the term ``ring'' to refer to their non-overlapping portion: for example, ring \#2 contains all words that are in~$\mathcal{L}_2$ but not in~$\mathcal{L}_1$ (see Table~\ref{tab:notation-semantic} for the general formula). 
For the sake of example, let us focus on the second cluster identified by Mean Shift: cluster \#2 corresponds to ring \#2 in the ego network, and the union of ring \#1 and ring \#2 corresponds to the 2nd layer of the ego network.
Another typical metric that is analysed in the context of social cognitive constraints is the scaling ratio~$\rho_i$ between layers $i$ and $i-1$, which, as discussed earlier, corresponds to the ratio between the size of consecutive layers (see Table~\ref{tab:notation-semantic} for its formula). The scaling ratio is an important measure of regularity, as it captures a relative pattern across layers, beyond the absolute values of their size. Taken together, the optimal number of layers $\tau^{(e)}$, the circle $\mathcal{L}_i^{(e)}$, and the scaling ratio $\rho_i^{(e)}$ fully characterise the ego network $e$.

\subsection{Results}
\label{sec:structural-results}

Here we study the ego networks of words in our four datasets, following the methodology described above.

\begin{figure}[h]
  \centering
  \iftoggle{NOFIG}{}{\includegraphics[width=\linewidth]{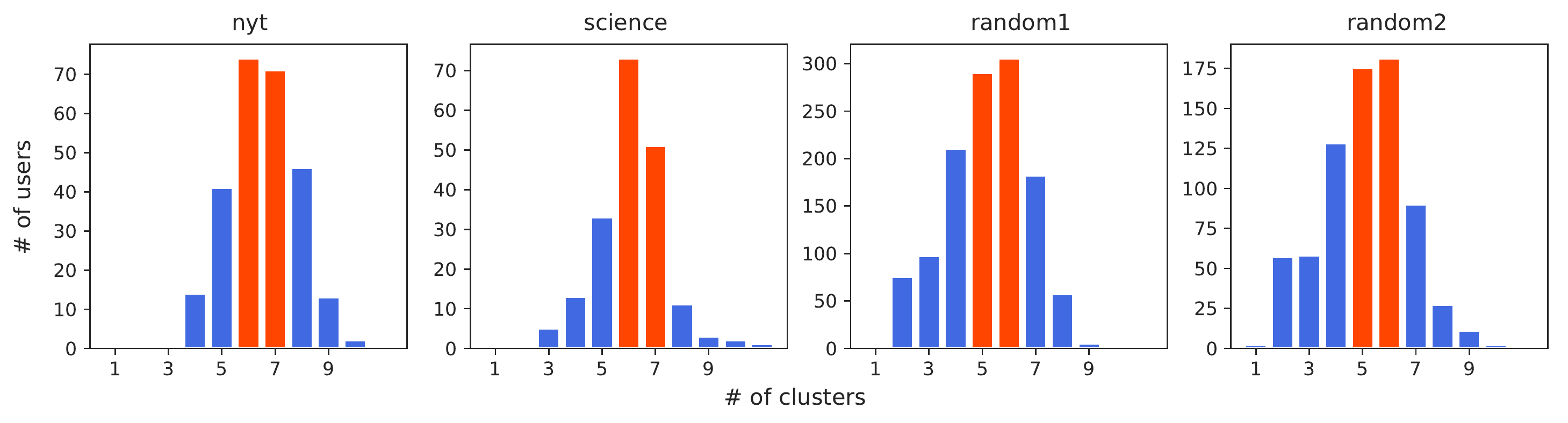}}
  \caption{{\bf Optimal number of clusters.} The clusters are obtained by applying Mean Shift to log-transformed frequencies. The most frequent number of clusters is highlighted in red. 
  \label{fig:clusters-hist}}
\end{figure}

The histograms of the obtained optimal number of layers $\tau$ are shown in Fig.~\ref{fig:clusters-hist}. It is interesting to note that, despite the heterogeneity of users (in terms of tweeting frequency), the distributions are always quite narrow, with peaks appearing consistently between 5 and 7 clusters.
Similarly to the social constraints case, also for language production, we observe a fairly regular and consistent structure. This is the first important result of the paper, hinting at the existence of structural invariants in cognitive processes.

\begin{figure}[h]
  \centering
  \iftoggle{NOFIG}{}{\includegraphics[width=\linewidth]{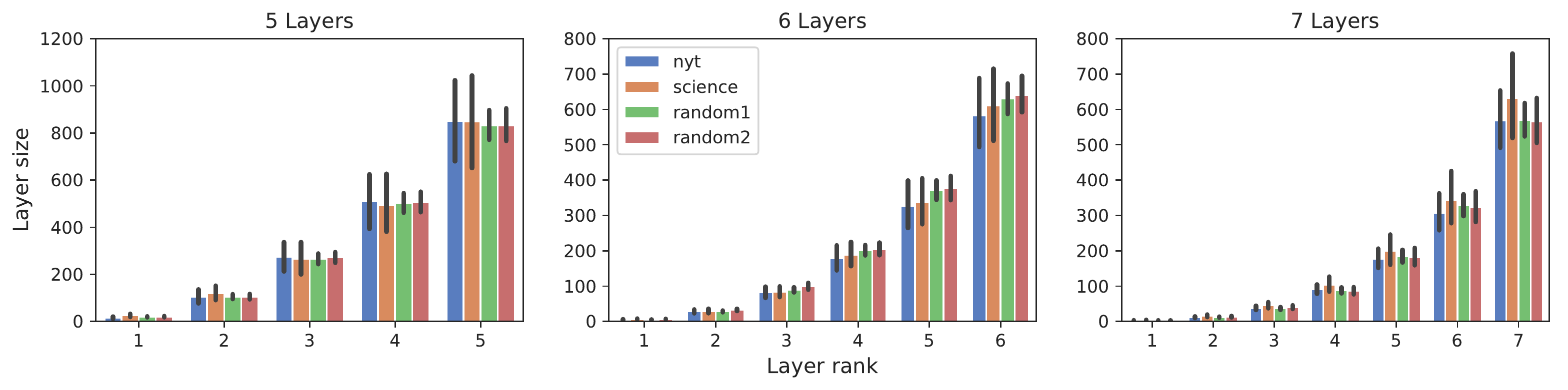}}
  \caption{{\bf Average layer size.} Each panel captures egos with a different optimal number of clusters. Error bars correspond to the 95\% confidence intervals.}
  \label{fig:cumul-size}
\end{figure}

We now study the size of the layers identified in Fig.~\ref{fig:clusters-hist}. For the sake of statistical reliability, we only consider those users whose optimal number of layers (as identified by Mean Shift) corresponds to the most popular number of layers (red bars) in Fig.~\ref{fig:clusters-hist}. This allows us to have a sufficient number of samples in each class. 
Fig.~\ref{fig:cumul-size} shows the average layer sizes for every dataset. For a given number of clusters, we observe again a striking regularity across the datasets, meaning that each layer has approximately the same size regardless of the category of users.

\begin{figure}[h]
  \centering
  \iftoggle{NOFIG}{}{\includegraphics[width=\linewidth]{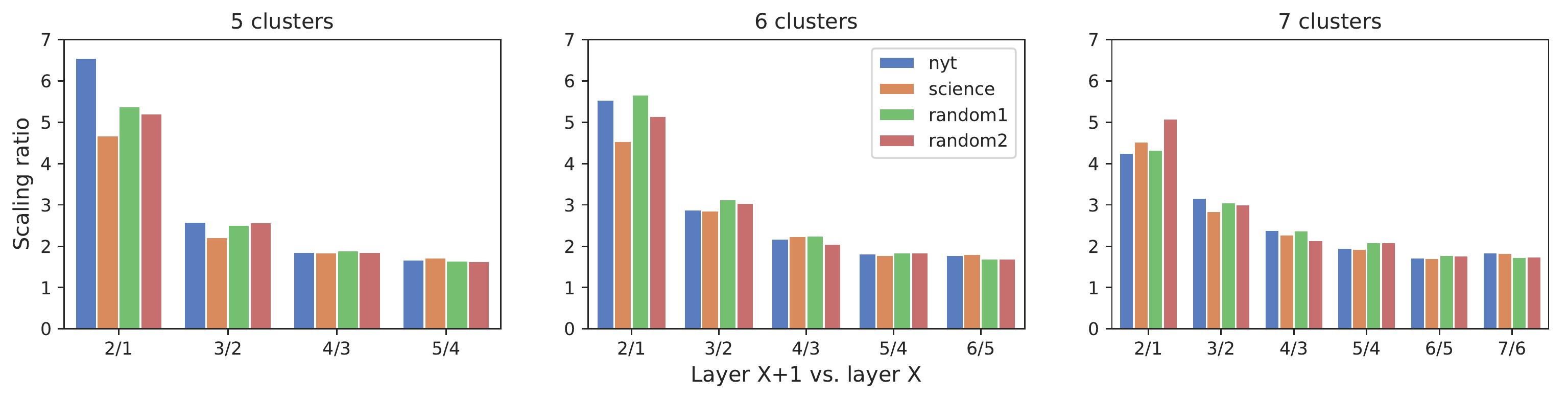}}
  \caption{{\bf Scaling ratio.} Each panel captures egos with a different optimal number of clusters. Error bars correspond to the 95\% confidence intervals.}
  \label{fig:scaling-ratio}
\end{figure}

Fig.~\ref{fig:scaling-ratio} shows the scaling ratio of the layers in language production. We can observe the following general behavior: the scaling ratio starts with a high value between layers \#1 and \#2, but always gets closer to 2-3 as we move outwards. This empirical rule is valid whatever the dataset (and whatever the observation period~\cite{ollivier2020}). This is another significant structural regularity, quite similar to the one found for social ego networks, as a further hint of cognitive constraints behind the way humans organise the words they use.

\begin{figure}[h]
  \centering
  \iftoggle{NOFIG}{}{\includegraphics[width=\linewidth]{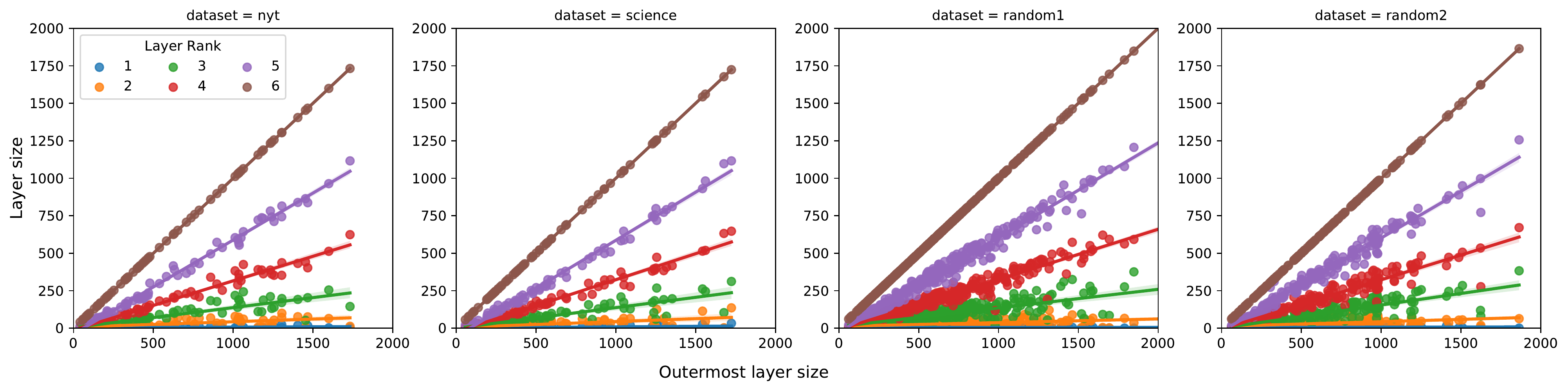}}
  \caption{{\bf Size of external layer vs individual layer size: linear regression plots.} The x-axis corresponds to the total number of unique words used by each user (corresponding to the size of the outermost layer), the y-axis to the individual layer sizes.}
  \label{fig:correl-cumul-size}
\end{figure}

In order to further investigate the structure of the word clusters, we compute the linear regression coefficients between the total number of unique words used by each user (corresponding to the size of the outermost layer) and the individual layer sizes. Due to space limits, in Table~\ref{table:coeff} we only report the exact coefficients for the journalists' dataset (but analogous results are obtained for the other categories) and in Fig.~\ref{fig:correl-cumul-size} we plot the linear regression for all the user categories. Note that the size of the most external cluster is basically the total number of words used by an individual in the observation window. It is thus interesting to see what happens when this number increases, i.e., if users who use more words distribute them uniformly across the clusters, or not. Table~\ref{table:coeff} shows two interesting features. First, it shows another regularity, as the size of all layers linearly increases with the most external cluster size, with the exception of the first one (Fig.~\ref{fig:correl-cumul-size}). Moreover, it is quite interesting to observe that the second-last and third-last layers consistently account for approximately 60\% and 30\% of the used words, irrespective of the number of clusters. This indicates that users with more clusters split, at a finer granularity, words used at the highest frequencies, i.e., they organise differently their innermost clusters, without modifying significantly the size of the most external ones. 

As a final comment on Fig.~\ref{fig:scaling-ratio}, please note that the innermost layer tends to be approximately five times smaller than the next one. This suggests that this layer, containing the most used words, may be drastically different from the others (as also evident from Table~\ref{table:coeff}). The characterization of this special layer will be the main focus of the next section. 

\begin{table}[!ht]
\centering
\caption{{\bf Size of external layer vs individual layer size: regression coefficients.} We report the linear regression coefficients obtained for the journalists dataset with~$T=1$ year.}
\scriptsize
\begin{tabular}{c c c c c c c c }
\toprule
\multirow{2}{*}{Opt. \# of clusters  } & \multicolumn{7}{c}{Cluster Rank}           \\ \cmidrule(r){2-8} 
                                    & 1    & 2    & 3    & 4    & 5    & 6    & 7 \\ \midrule
5 clusters                          & 0.02 & \cellcolor{gray!10}0.13 & \cellcolor{gray!25} 0.33 & \cellcolor{gray!50} 0.62 & 1.00    &      &   \\ 
6 clusters                          & 0.01 & 0.04 & \cellcolor{gray!10}0.14 & \cellcolor{gray!25} 0.32 & \cellcolor{gray!50} 0.59 & 1.00    &   \\ 
7 clusters                          & 0.00 & 0.02 & 0.06 & \cellcolor{gray!10}0.16 & \cellcolor{gray!25} 0.32 & \cellcolor{gray!50} 0.56 & 1.00 \\ \bottomrule
\end{tabular}
\label{table:coeff}
\end{table}

\subsection{Discussion}
\label{sec:structural-results-discussion}

We summarise below the main  results of the section.
\vspace{-5pt}
\begin{itemize}
    \item Individual distributions of word frequencies are divided into a consistent number of groups. Since word frequencies impact the cognitive processes underlying word learning and retrieval in the mental lexicon~\cite{perfetti2005word}, these groups can be an indirect trace of these processes' properties. The number of groups is only marginally affected by the class (specialized or generic) the users belong.
    \item Structural invariants in terms of layer sizes and scaling ratio are observed, similarly to the well-known results from the social domain~\cite{Dunbar2015}. Specifically, we found that the size of the layers approximately doubles when moving from layer~$i$ to layer~$i + 1$, with the only exception of the first layer.
    \item Users with more layers organise differently their innermost layer, without modifying significantly the size of the most external ones, which consistently account for approximately 60\% and 30\% of the used words, irrespective of the number of clusters of the user. 
\end{itemize}

\section{Semantic analysis of the ego network of words}
\label{sec:semantic}

We have treated words as simple tokens so far. However, words have meanings and they can be linked to specific topics.
In this section, we want to go beyond words and investigate which topics they refer to and how they are distributed in the different rings of the ego network. The analysis of this section revolves around the concept of \emph{semantic profile} of a ring (in the ego network of words), which captures the topics associated with the words in the ring. 
Once semantic profiles are obtained, we are able to address the following high-level question: are all rings similar in the topics they contain, or does the ego network organize the topics in its rings in a specific way?

For the convenience of the reader, we summarise in Table~\ref{tab:notation-semantic} the notation used throughout the section.

\begin{table}[!ht]
\centering
\caption{{\bf Summary of the notation used in the semantic analysis}}
\begin{tabular}{@{}lp{0.75\textwidth}@{}}
\toprule
 \textbf{Symbol} & \textbf{Description} \\ \midrule
$\mathcal{E}$  & Set of all ego networks $\mathcal{E}$\\
$e \in \mathcal{E}$  & Ego networks $e$ belonging to the set of all ego networks $\mathcal{E}$\\
$c \in \mathcal{C}$     & Topic $c$ belonging to the set of all topics $\mathcal{C}$\\
$m \in \mathcal{T}$  & Tweet $m$ belonging to the set of all tweets $\mathcal{T}$\\
$P_m$ & Semantic profile of tweet $m$, according to HDBSCAN \\ $P_m(c)$ & Likelihood that tweet $m$ belongs to topic $c$, according to the semantic profile of the tweet\\
$\mathcal{W}(e,r)$      & Set of non-distinct words in ring $r$ of ego network $e$ \\
$\mathcal{W}_u(e,r)$    & Set of distinct words in ring $r$ of ego network $e$ \\
$\mathcal{W}(e,w_u)$ & Set of occurrences of the unique word $w_u$ in the ego network $e$ \\
$O(e,r)$ & Number of word occurrences in ring $r$ of ego network $e$\\
$o(w_u,e)$ & Number of occurrences associated with the unique word $w_u$ of ego network $e$\\
$P_r^{(e)}$   & Semantic profile of ring $r$ of ego network $e$ \\
$P_r^{(e)}(c)$& Probability of observing topic $c$ in $P_r^{(e)}$ of ring $r$ in ego network $e$ \\
$P_{w_u}^{(e)}$& Topic distribution of unique word $w_u$ in ego network $e$ \\
$P_{w_u}^{(e)}(c)$& Probability of observing topic $c$ in $P_{w_u}^{(e)}$ for $w_u$ in ego network $e$ \\
$\mathcal{N}(e,r)$      & The number of topics discussed in ring $r$ of ego network $e$ \\
$\mathcal{N}_{norm}(e,r)$& $\mathcal{N}(e,r)$ normalised by the total number of word occurrences in $r$\\
$H(e,r)$                & Entropy of the semantic profile $P_r^{(e)}$\\
$\delta_{JS}\left(P_{r_i}^{(e)} || P_{r_j}^{(e)}\right)$ & Distance between the semantic profiles of rings $i$ and $j$\\ 
$U_r^{(e)}$ & Set of primary topics for ring $r$ of ego network $e$\\
$L_r^{(e)}$ & Set of non-primary topics for ring $r$ of ego network $e$\\
$K_{TOP(r_x)}^{r_y}$ & Coverage of $r_x$'s primary topics in $r_y$'s semantic profile\\
$S_{TOP(r_x)}^{r_y}$ & Strength of $r_x$'s primary topics in $r_y$'s semantic profile\\
$S_{BOTTOM}^{r_y}$ & Strength of $r_x$'s non-primary topics in $r_y$'s semantic profile\\
$S_{TOP(r_x, r_y)}^{r_y}$ & Strength of topics that are primary for both $r_x$ and $r_y$ in $r_y$'s semantic profile\\
$S_{TOP(r_x),BOTTOM(r_y}^{r_y})$ & Strength of topics that are primary for $r_x$ but not for $r_y$ in $r_y$'s semantic profile\\
$\sigma_{TOP(r_x, r_y)}^{r_y}$ & Strength of topics that are primary for both $r_x$ and $r_y$ with respect to the average strength of primary topics in $r_y$'s semantic profile\\
$\sigma_{TOP(r_x),BOTTOM(r_y}^{r_y}$ & Strength of topics that are primary for $r_x$ but not for $r_y$ with respect to the average strength of non-primary topics in $r_y$'s semantic profile\\
\bottomrule
\end{tabular}
\label{tab:notation-semantic}
\end{table}

\subsection{How to build semantic profiles}
\label{sec:semantic-methodology}

In this section, we describe how we carry out the semantic analysis of the ego network of words. First, in Section~\ref{sec:topic-preliminaries}, we motivate our selection of the BERTopic framework for topic extraction. Then, in Section~\ref{sec:topic-extract}, we illustrate the steps for topic extraction. At the end of this process, each word occurrence in the ego network is associated with a specific topic. Accounting for the popularity of each topic in the rings of the ego network, in Section~\ref{sec:semprof-extraction} we build the \emph{semantic profile} of the ego network ring, as the topic distribution of the words in that ring. 

\subsubsection{Preliminaries}
\label{sec:topic-preliminaries}

To calculate a semantic profile, we choose to consider the meaning of each word in its context rather than using a semantic dictionary~\cite{SEMCAT} (a dataset where each word is mapped to a semantic category), which would not be able to detect more complex topics  and would miss some meanings for a polysemous word. 
We acknowledge that a lot of effort has been put in the direction of ontologies in order to understand more precisely the interests of users, specifically on Twitter. Ontologies map knowledge of specific domains, such as Athena~\cite{frasincar2009semantic}, which is a semantic web database extracted from a news portal that can be used for news recommendation purposes~\cite{jonnalagedda2013personalized}, or the BBC ontologies extracted from the BBC  corpus of news, which allows politically-oriented topic mining~\cite{abu2018twitter}. However, even if their drawbacks (such as the rigidity of the knowledge model) can be partly fixed by coupling them with models based on embedding~\cite{mevznar2021link}, we prefer having the maximum freedom in the topic identification process by using a transformers-based model such as BERT~\cite{devlin2018bert} which is the current state of the art in text embedding and then using an unsupervised method to detect topics.

\subsubsection{Extraction of the topics}
\label{sec:topic-extract}

\begin{figure}[h]
  \centering
  \iftoggle{NOFIG}{}{\includegraphics[width=\linewidth]{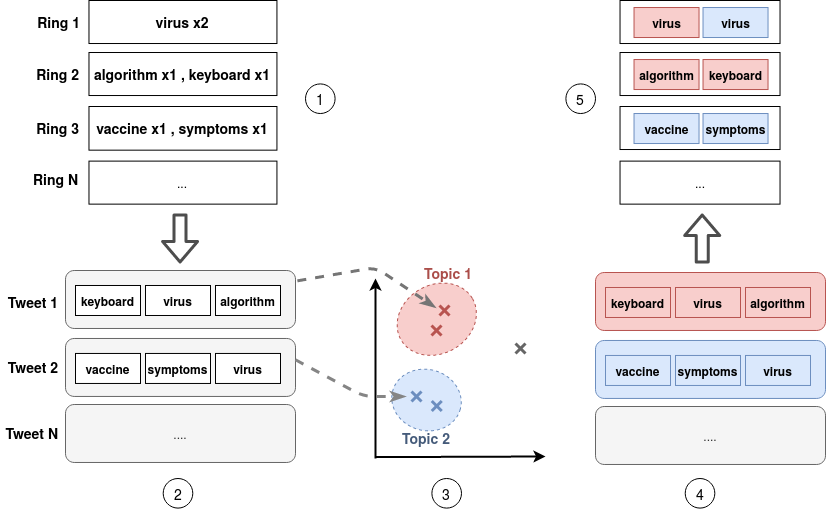}}
  \caption{{\bf Obtaining the semantic profile of the rings of an ego network.} (1) The ego network's rings organize a user's vocabulary based on the frequencies of the words. (2) For a given word, its occurrences in the user timeline are coming most likely from different tweets. (3) The tweets are classified by topic thanks to the BERTopic framework. (4) Each word occurrence is assigned the very same topic as the tweets it belongs to. (5) If we consider a ring as a multiset of words (with repetitions) the semantic profile is the distribution of the topics among those words.}
  \label{fig:full-process-bertopic}
\end{figure}

In order to avoid some issues with polysemous words, we must consider the ring of an ego network not only as a set of single words associated with a frequency of use but as a set of words with a given number of occurrences (from which the frequency is derived), each occurrence belonging to a user's tweet.
We aim to associate each word occurrence with a topic. We first classify (in an unsupervised way) the tweets by topic using the BERTopic framework~\cite{grootendorst2020bertopic}, then all word occurrences that constitute a tweet are assigned the same topic as the tweet itself (Fig.~\ref{fig:full-process-bertopic}).

For the current analysis, we chose to focus only on ego networks with six rings, the case covering the most users. As described in the following, the BERTopic framework uses sequentially BERT~\cite{devlin2018bert} for tweet embedding, UMAP~\cite{mcinnes2018umap} for dimension reduction, and HDBSCAN~\cite{McInnes2017} for clustering those tweet embeddings in a low-dimensional subspace. 

\paragraph{Tweet embedding with BERT.}
BERT~\cite{devlin2018bert}, which achieves state-of-the-art performance for natural language understanding, is used to assign to each tweet a point in the embedding space which is supposed to be a vector representation of its semantic meaning. BERT is a bidirectional transformer developed by Google, trained on the BookCorpus~\cite{zhu2015aligning} and Wikipedia in English. It, therefore, relies on all the linguistic knowledge learned from a very large corpus to perform this task. 
BERT yields topics along 768 dimensions. 

\paragraph{Dimensionality reduction with UMAP.} In order to mitigate the curse of dimensionality (to which clustering algorithm based on k-nearest neighbors are particularly sensible~\cite{radovanovic2010hubs}), we use the UMAP clustering algorithm (with settings \texttt{n_neighbors=15, n_components=5, metric='cosine'} and the python package \texttt{umap v0.1.1}) to reduce the embedding space down to five dimensions as recommended in the BERTopic framework~\cite{grootendorst2020bertopic}. UMAP, like the T-SNE~\cite{van2008visualizing} algorithm, is able to capture latent non-linear dimensions but in a more scalable way. 

\begin{figure}[!h]
  \centering
 \iftoggle{NOFIG}{}{\includegraphics[width=\linewidth]{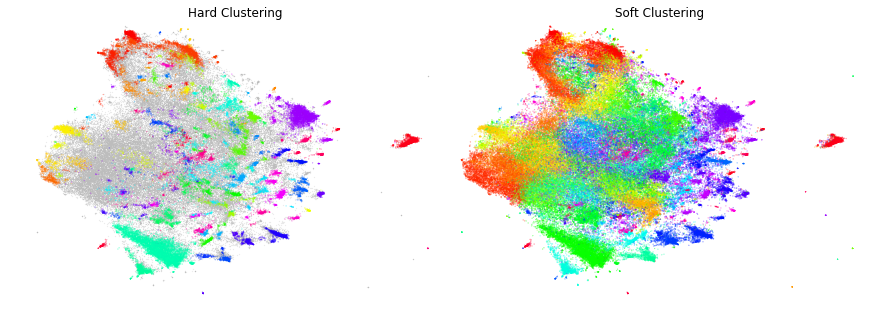}}
  \caption{{\bf 2D visualization of the HDBSCAN results on the Journalists dataset with both hard and soft clustering.} 265 clusters are found (they are the same in both cases). In the first case, each point is classified as either belonging to a single cluster (colored points) or as an outlier (grey point), whereas in the second case each point is assigned a likelihood to belong to each cluster (the points take the color of the cluster they belong to most likely).}
  \label{fig:hardsoft-2d}
\end{figure}

\paragraph{HDBSCAN for clustering topics.} HDBSCAN~\cite{McInnes2017} is also able to find non-linear cluster structures from the density, as well as outliers, like DBSCAN (Fig.~\ref{fig:hardsoft-2d}). However, instead of deciding the contours of a cluster based on a fixed density threshold, HDBSCAN uses hierarchical clustering (single linkage) to find the most stable partition. Here we use HDBSCAN with following settings: \texttt{min_cluster_size=15, metric='euclidean', cluster_selection_method='eom', prediction_data=True} with the python package \texttt{hdbscan v0.8.26}. Thanks to BERT embedding, the clusters of tweets we obtain are semantically homogeneous, and therefore represent the dominant topics of the dataset. Under these conditions, we can consider that a cluster corresponds to a topic. 

Table~\ref{tab:tweets-cluster} shows the percentage of outliers detected by HDBSCAN, which corresponds to the percentage of tweets that cannot be associated with a specific topic. Since this percentage is quite high, even with the most conservative configurations (with the least outliers), we also assess the cluster configuration (\textit{i.e.}, the topic assignment) induced by a soft clustering approach. Indeed HDBSCAN allows two types of clustering: hard clustering, which classifies each tweet in one and only one cluster (or as an outlier), and soft clustering, which is able to measure the proximity of a tweet to several different clusters. The advantage is that it is possible to obtain this proximity even for outliers, which allows us to integrate them into the analysis.
When using it for soft clustering, HDBSCAN provides, for each point (tweet)~$m$, a probability distribution $P_m$ such that $P_m(c)$ is the likelihood that this point belongs to the cluster (topic) $c$, with $\sum_{c \in \mathcal{C}} P_m(c) \leq 1$ ($\mathcal{C}$ being the set of topics). Thus, with soft clustering, the tweet is not assigned a single topic but a probability distribution over all the topics.
For clarity reasons, in the case of hard clustering - where the tweet~$m$ is directly assigned one topic~$c_m$ - let us use the same notation $P_m$, where $P_m(c_m)$ is equal to 1 and zero otherwise.
We will use these two configurations (hard clustering and soft clustering) to build two separate semantic profiles for each ego network ring. 
In~\nameref{S1-Appendix} we discuss in detail why hard clustering is better suited for our analysis.

\begin{table}[!ht]
\centering
\caption{{\bf Topics per dataset.} Each topic corresponds to a cluster identified by HDBSCAN.}
\scriptsize
\setlength{\tabcolsep}{0.5em}
\renewcommand{\arraystretch}{1.2}
    \begin{tabular}{lcc}
        \toprule
            Datasets & Number of topics & \% of outliers \\  \midrule
            NYT Journalists & 265 & 69.3\% \\
            Science Writers & 223 & 71.8\% \\
            Random Users \#1 & 2940 & 68.6\% \\
            Random Users \#2 & 2577 & 70.0\% \\
        \bottomrule
    \end{tabular}
    \label{tab:tweets-cluster}
\end{table}

\begin{figure}[!h]
  \centering
  \iftoggle{NOFIG}{}{\includegraphics[width=\linewidth]{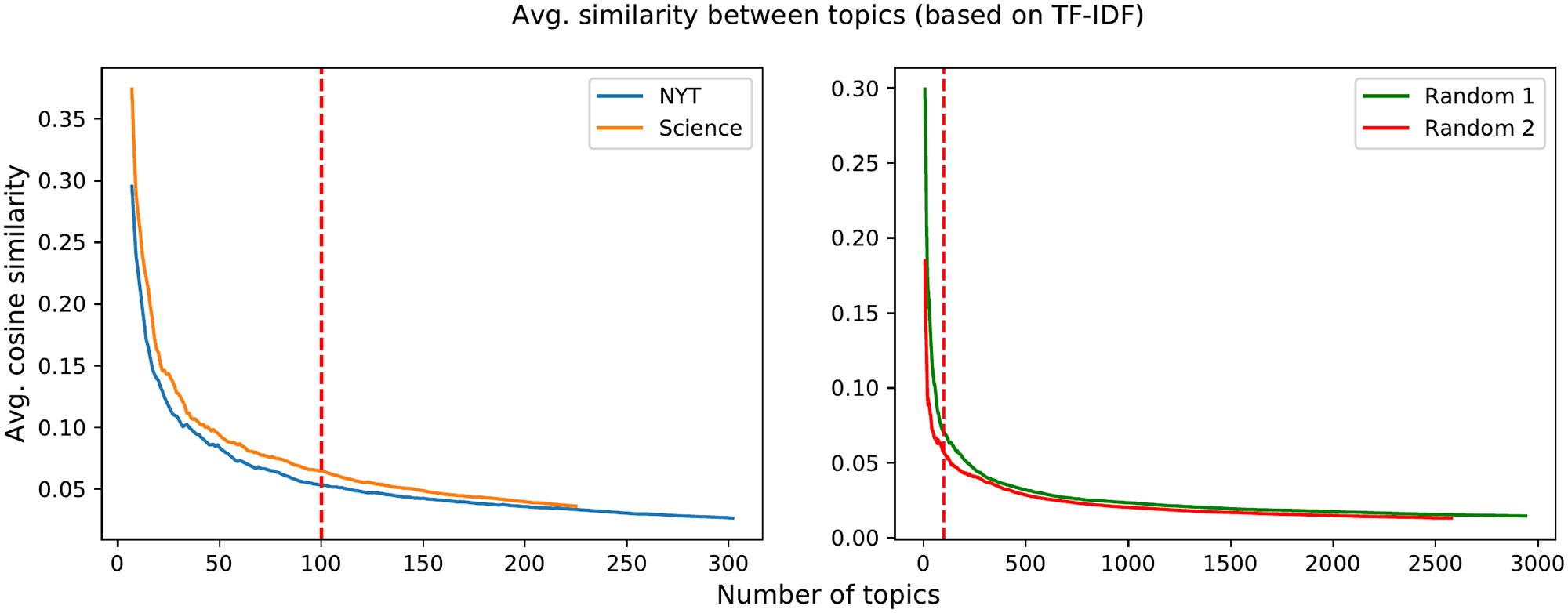}}
  \caption{{\bf Number of topics vs. average topic similarity.} The threshold of one hundred topics is marked with the dashed red line. This threshold is situated at the end of the bend for specialized datasets, and in the middle of the bend for both random datasets.}
  \label{fig:topic-merge-sim}
\end{figure}

\paragraph{Reduction of the number of topics} As shown in Table~\ref{tab:tweets-cluster}, the different datasets feature a different number of topics. In order to be able to compare the datasets, we reduced the number of topics down to the same number of topics (this set of topics - which is different for each dataset - will be noted as $\mathcal{C}$ from now on). Let us denote with $\mathcal{C}'$ the full set of topics. Our goal is to merge them together until we obtain the target number of topics. To do so, the following operation is repeated: merge the smallest cluster $c_1'$ (in the hard clustered configuration) with the cluster $c_2'$ to which $c_1'$ is semantically the closest. This semantic similarity is calculated as follows: all the tweets are grouped in a single document by cluster, then a TF-IDF vector is calculated for each of them. The similarity between the two topics is the cosine of their TF-IDF representation. The probability of the new topic $c_1' \cup c_2'$ is accordingly updated, for each tweet $m$, as $P_m(c_1' \cup c_2') = P_m(c_1') + P_m(c_2')$. When merging step by step the clusters, the average similarity between them increases as can be seen in Figure~\ref{fig:topic-merge-sim}. In the case of journalists and science writers, we see that exceeding 100 topics no longer allows the emergence of topics that are radically different from the others, while still enabling an acceptable number of topics to be isolated. Thus, in order to be able to compare the results related to the different datasets, we have chosen to limit the number of topics to 100 for each of them. For the sake of comparison, the 100 topics obtained for the hard clustering configuration are also used for topic reduction in the soft clustering case.
%
%
%
This operation allows us to narrow down to one hundred topics the different semantic fields addressed in the same dataset while trying to provoke the least changes in the topic reassignment. 



\subsubsection{Extraction of the semantic profile}
\label{sec:semprof-extraction}

We define the semantic profile of an ego network ring as the distribution of topics to which the word occurrences that the ring contains (multiple occurrences of the same word may come from different contexts and thus refer to different topics) belong.  Note that this analysis is carried out at the ring level, and not the circle level because circles are concentric and cumulative, thus the semantic profiles of circles would include by default overlapping topics, hence creating a bias in the analysis (similarly to counting topics twice).
After the preprocessing described in the previous section, each word occurrence is associated with a topic (or several, in the soft clustered case), thus we can compute for each ego network’s ring a topic distribution based on the word occurrences it contains.

\begin{figure}[h]
  \centering
 \iftoggle{NOFIG}{}{ \includegraphics[width=0.6\linewidth]{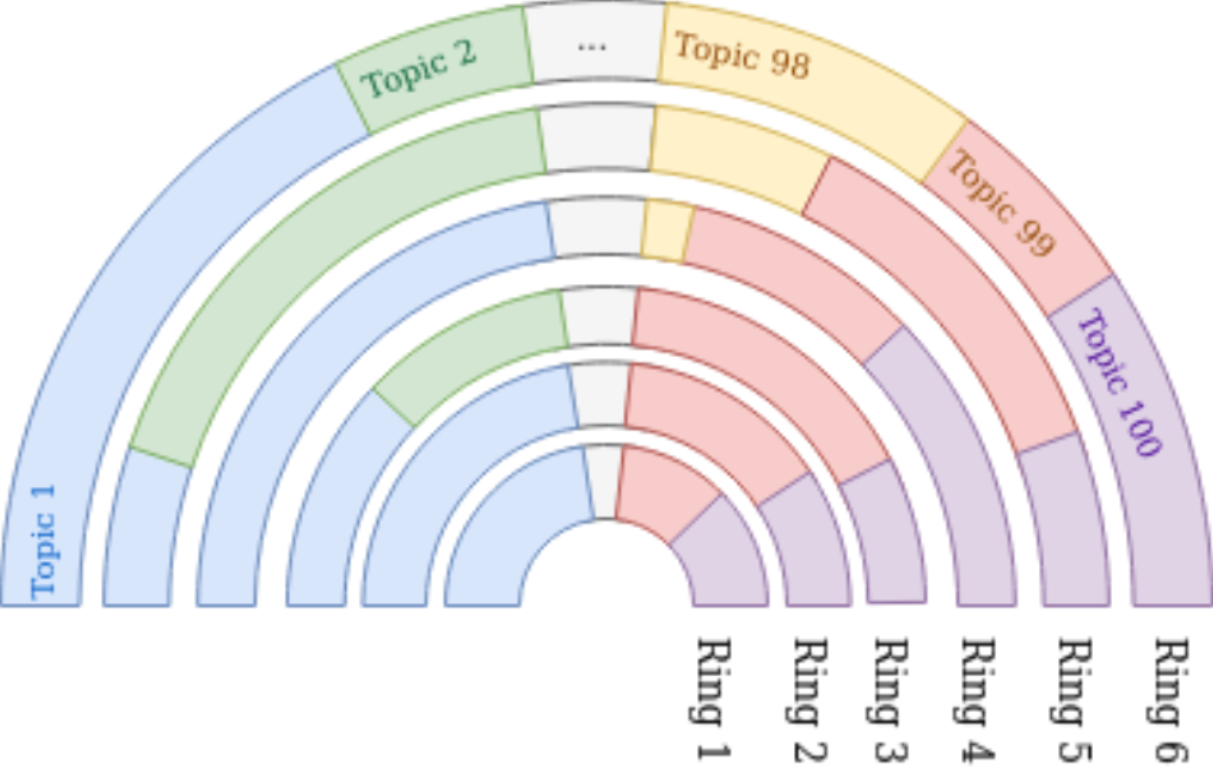}}
  \caption{{\bf Semantic profile illustration}. Each ring is associated with a topic distribution.}
  \label{fig:semantic-profile}
\end{figure}

Let $\mathcal{W}(e,r)$ be the set of word occurrences contained in ring $r$ of the ego network $e$, and $m(w)$ the tweet the word occurrence $w$ belongs to. The probability $P_r^{(e)}(c)$ of observing topic $c$ in ring $r$ of ego network $e$ is defined as follows:
\begin{equation}
P_r^{(e)}(c) =  \frac{\sum_{w \in \mathcal{W}(e,r)} P_{m(w)}(c)}{\sum_{c \in \mathcal{C}} \sum_{w \in \mathcal{W}(e,r)} P_{m(w)}(c)},
\end{equation}
where $\sum_{c \in \mathcal{C}} P_r^{(e)}(c) = 1$. More in general, we denote with $P_r^{(e)}$ the semantic profile of ring $r$ in ego network $e$ (depicted in Fig.~\ref{fig:semantic-profile}). For this reason, we will also refer to $P_r^{(e)}(c)$ as the share of $c$ in the semantic profile $P_r^{(e)}$ of $r$
This unique semantic profile will be the starting point for all subsequent analyses in this section. In~\nameref{S1-Appendix}, we provide four tables (one for each dataset) that detail for every topic the most characteristic words and the average share in the rings.

\noindent
\emph{Note:} Two different semantic profiles can be built, depending on whether topics are assigned using hard vs soft clustering. 
In~\nameref{S1-Appendix}
we show that the use of soft clustering (and thus the inclusion of outliers) does not improve the reliability of the analysis. It gives too much importance to noisy data which favors the emergence of very generalized "super topics" that dominate all semantic profiles. We, therefore, present in Section~\ref{sec:results-topics} only the results obtained with hard clustering. 
In~\nameref{S1-Appendix}
we discuss soft versus hard clustering in detail and motivate why hard clustering is better suited for our analysis.

\subsection{Metrics for the analysis of semantic profiles}
\label{sec:sempro-metrics}

After following the steps described in Section~\ref{sec:semantic-methodology}, we end up with a semantic profile for each ring of an ego network. In the following we discuss (i) how to characterise individual semantic profiles (Section~\ref{sec:sempro-charact}), (ii) how to compare semantic profiles (Section~\ref{sec:sempro-ring-comparison}), and (iii) how to leverage semantic profiles to investigate the role of the most important topics (Section~\ref{sec:important-topics}). 

\subsubsection{Characterization of the semantic profile}
\label{sec:sempro-charact}

Let us consider a ring $r$ of ego network $e$ for which we have extracted the semantic profile as discussed above. The semantic profile tells us how many distinct topics the words in ring $r$ touch upon. Formally, the number of topics associated with a given ring can be calculated as follows:
\begin{equation}
	\mathcal{N}(e,r) = \sum_{c \in \mathcal{C}} \mathbbm{1}_{P_r^{(e)}(c) > 0},
\end{equation}
where we denoted with $P_r^{(e)}(c)$ the probability of a observing topic $c$ in the semantic profile $P_r^{(e)}$ of ring $r$, and $\mathbbm{1}$ is the indicator function.
Note, though, that $\mathcal{N}(e,r)$ may offer only a partial perspective. In fact, rings have very different sizes (as discussed in Section~\ref{sec:structural}) and it is expected to be much easier for larger rings (i.e., rings containing many words) to span a larger range of topics. For this reason, we will compare $\mathcal{N}(e,r)$ with its normalised version:
\begin{equation}
    \mathcal{N}_{norm}(e,r) = \frac{\mathcal{N}(e,r)}{|\mathcal{W}(e,r)|},
\end{equation}
where we weigh the number of topics ``generated'' by the ring by the number of word occurrences contained in the ring (denoted with $|\mathcal{W}(e,r)|$).


$\mathcal{N}(e,r)$ and $\mathcal{N}_{norm}(e,r)$ account for the mere presence of topics, regardless of their frequency of use. To capture the latter dimension, we next measure the entropy of $P_r^{(e)}$. Recalling that $P_r^{(e)}$ is in fact a probability distribution, its Shannon entropy reflects its diversity: the entropy (and diversity) is maximum if a ring contains all topics equally (i.e., with the same values of $P_r^{(e)}(c)$), while the entropy is minimum if a ring contains only one topic. So, the greater the entropy, the greater the diversity. Denoting  with $H(e,r)$ the entropy of the ring $r$ in ego $e$, its definition is as follows:
\begin{equation}
	H(e,r) = - \sum_{c \in \mathcal{C}} P_r^{(e)}(c) \times \log\left(P_r^{(e)}(c)\right).
\end{equation}
For the 100 topics we consider, the minimum entropy is 0 and the maximum entropy is about~4.60.

In Section~\ref{sec:results-topics}, the average of $\mathcal{N}(e,r)$, $\mathcal{N}_{norm}(e,r)$, and $H(e,r)$ across all ego networks will be presented, i.e., $\mathcal{N}(r) = \frac{1}{|\mathcal{E}|} \sum_{e \in \mathcal{E}} \mathcal{N}(e,r)$ (analogously for the others).

\subsubsection{Comparing the semantic profiles of different rings}
\label{sec:sempro-ring-comparison}

Once we know which topics are covered by each ring of an ego network, the first step is to find out whether their semantic profile differs from one ring to another one or, instead, if the distribution is homogeneous over the whole ego network.
Since all semantic profiles are based on the same 100 topics, it is easy to obtain a distance measure to compare the rings with one another. Recalling that the semantic profile is a probability distribution, for this purpose we can use the Jensen-Shannon (JS) divergence~\cite{lin1991divergence}, which allows us to calculate the proximity between the 100-topic distributions that we obtained previously. Then, the corresponding JS distance is conventionally obtained as the square root of the JS divergence~\cite{osterreicher2003new}. The JS divergence is basically a symmetric version of the well-known Kullblack-Leibler (KL) divergence, which is a standard metric for capturing the distance between probability distributions. For a tagged ego $e$, the KL divergence $D_{KL}$ between two semantic profiles $P_{r_i}^{(e)}$ and $P_{r_j}^{(e)}$ of rings $i$ and $j$ for ego network $e$ can be computed as follows:
\begin{equation}
    D_{KL}\left(P_{r_i}^{(e)} || P_{r_j}^{(e)} \right) = \sum_{c \in \mathcal{C}} P_{r_i}^{(e)}(c) \times log \left(\frac{P_{r_i}^{(e)}(c)}{P_{r_j}^{(e)}(c)}\right).
\end{equation}
From $D_{KL}(P_{r_i}^{(e)} || P_{r_j}^{(e)})$, the JS divergence can be obtained as:
\begin{equation}
    D_{JS}\left(P_{r_i}^{(e)} || P_{r_j}^{(e)}\right) = \frac{D_{KL}\left(P_{r_i}^{(e)}||M\right) + D_{KL}\left(P_{r_j}^{(e)}||M\right)}{2},
\end{equation}
with $M = \frac{P_{r_i}^{(e)} + P_{r_j}^{(e)}}{2}$. Then we go from divergence $D$ to distance $
\delta$ by taking the square root: $\delta_{JS}(P_{r_i}^{(e)}, P_{r_j}^{(e)}) = \sqrt{D_{JS}\left(P_{r_i}^{(e)} || P_{r_j}^{(e)}\right)}$. Note that the JS distance is bounded as $0 \leq \delta_{JS}\left(P_{r_i}^{(e)} || P_{r_j}^{(e)}\right) \leq \sqrt{log(2)} \approx 0.83$.

Once we have obtained a $\delta_{JS}\left(P_{r_i}, P_{r_j}\right)$, we compute its average across all ego networks in a standard way, i.e., $\delta_{JS}^{(e)}\left(P_{r_i}, P_{r_j}\right) = \frac{1}{|\mathcal{E}|} \sum_{e \in \mathcal{E}} \delta_{JS}^{(e)}\left(P_{r_i}, P_{r_j}\right)$

\subsubsection{Capturing important topics and their cross-rings effects}
\label{sec:important-topics}

Given a semantic profile $P_r^{(e)}$, we can check whether some topics are more important than others, and, if this is the case, whether they play a special role in the ego network's rings. We consider whether topics can be divided in two classes, \textit{i.e.}, ``important" and ``not-important" topics for each ring. To do so, we cluster the topics according to their presence in the specific ring under study, i.e, according to the values of $P_r^{(e)}(c)$ where $c \in \mathcal{C}$.
To this aim, we use the Jenks  algorithm~\cite{jenks1967data} which allows finding natural breaks in the frequency distribution (similarly to k-means, we have to specify~$k$, the number of groups we want to obtain). We rely on the Silhouette score~\cite{rousseeuw1987silhouettes} to validate the clustering results. Since we just want to find one natural break that separates important topics from the others, we set $k=2$.  Words are split into two groups, one with high-frequency use, and the other with low-frequency use. The former is the set of important (or primary) topics referred to as $U^{(e)}_r$ (where $e$ is the ego network and $r$ is the ring number), and the latter is the set of non-important topics as $L^{(e)}_r$. 

Once we have obtained $U^{(e)}_r$ and $L^{(e)}_r$, for all ego networks and for all rings, we can investigate whether primary topics in one ring play a special role in other rings as well. Let us focus on two rings $x$ and $y$. We define $K_{TOP(r_x)}^{r_y}$ as the coverage of $r_x$'s primary topics in ring $r_y$. This metric captures the cumulative presence of $r_x$'s primary topics in $r_y$.
\begin{equation}
    K_{TOP(r_x)}^{r_y} = \frac{1}{|\mathcal{E}|} \sum_{e \in \mathcal{E}}  \sum_{c \in U^{(e)}_{r_x}} P_{r_y}^{(e)}(c).
\end{equation}
Then, to capture the average individual strength of $r_x$'s primary topics in $r_y$, we define a complementary metric $S_{TOP(r_x)}^{r_y}$ (with an averaging factor $\frac{1}{|U^{(e)}_{r_x}|}$) as follows:
\begin{equation}
    S_{TOP(r_x)}^{r_y} = \frac{1}{|\mathcal{E}|} \sum_{e \in \mathcal{E}} \frac{1}{|U^{(e)}_{r_x}|} \sum_{c \in U^{(e)}_{r_x}} P_{r_y}^{(e)}(c).
\end{equation}
Basically, $S_{TOP(r_x)}^{r_y}$ measures the average share of \emph{each} $r_x$'s primary topics in another ring of the same ego network. Similarly, we can compute $S_{BOTTOM(r_x)}^{r_y}$ by replacing $U^{(e)}_{r_x}$ with $L^{(e)}_{r_x}$ in the above equation.
This approach can be generalized to more complex cases. For example, we can study the strength of topics that are important in \emph{both} $r_x$ and $r_y$ in the semantic profile of ring $r_y$. This would be equivalent to the following:
\begin{equation}
S_{TOP(r_x,r_y)}^{r_y} = \frac{1}{|\mathcal{E}|} \sum_{e \in \mathcal{E}} \frac{1}{|U_{r_x}^{(e)} \cap U_{r_y}^{(e)}|} \sum_{c \in U_{r_x}^{(e)} \cap U_{r_y}^{(e)}} P_{r_y}^{(e)}(c).
\end{equation}
Analogously, we can study the opposite effect, i.e., what is the strength of topics that are important in $r_x$ but \emph{not} in $r_y$ in the semantic profile of $r_y$. In this case, the formula will be the following:
\begin{equation}
S_{TOP(r_x),BOTTOM(r_y)}^{r_y} = \frac{1}{|\mathcal{E}|} \sum_{e \in \mathcal{E}} \frac{1}{|U_{r_x}^{(e)} \cap L_{r_y}^{(e)}|} \sum_{c \in U_{r_x}^{(e)} \cap L_{r_y}^{(e)}} P_{r_y}^{(e)}(c).
\end{equation}
All the above metrics capture the \emph{pulling power} of ring $r_x$ on ring $r_y$. 

Another interesting perspective is whether topics that are primary elsewhere tend to be more or less dominant than the average topic in $U_{r_y}^{(e)}$ or $L_{r_x}^{(e)}$. This effect can be measured as follows:
\begin{equation}
    \sigma_{TOP(r_x,r_y)}^{r_y} = S_{TOP(r_x,r_y)}^{r_y} - S_{TOP(r_y)}^{r_y},
\end{equation}
where we basically compute the difference between the strength of topics that are primary in both $r_x$ and $r_y$ and the average strength of all primary topics in $r_y$. The complementary perspective is whether topics that are primary elsewhere tend to be more or less dominant than the average non-primary topic in ${r_y}$. To this aim, we leverage the following:
\begin{equation}
    \sigma_{TOP(r_x),BOTTOM(r_y)}^{r_y} = S_{TOP(r_x),BOTTOM(r_y)}^{r_y} - S_{BOTTOM(r_y)}^{r_y}.
\end{equation}
which follows the same line of reasoning as $\sigma_{TOP(r_x,r_y)}^{r_y}$.
\subsection{Results}
\label{sec:results-topics}

In this section, we study the semantic profiles in the ego networks of the Twitter users in our four datasets (Section~\ref{sec:dataset}).

\subsubsection{Ring \#1 is special in the ego networks of words}
\label{sec:results-topics-per-ring}

We start our analysis by studying how topics are associated with the different rings. For each ego network $e$, we will compute the number of topics per ring ($\mathcal{N}(e,r)$ and $\mathcal{N}_{norm}(e,r)$, its normalized version) and their entropy $H(e,r)$. These metrics are then averaged across all egos, as described in Section~\ref{sec:sempro-metrics}, and 95\% confidence intervals are shown.

\begin{figure}[H]
  \centering
  \iftoggle{NOFIG}{}{\includegraphics[width=\linewidth]{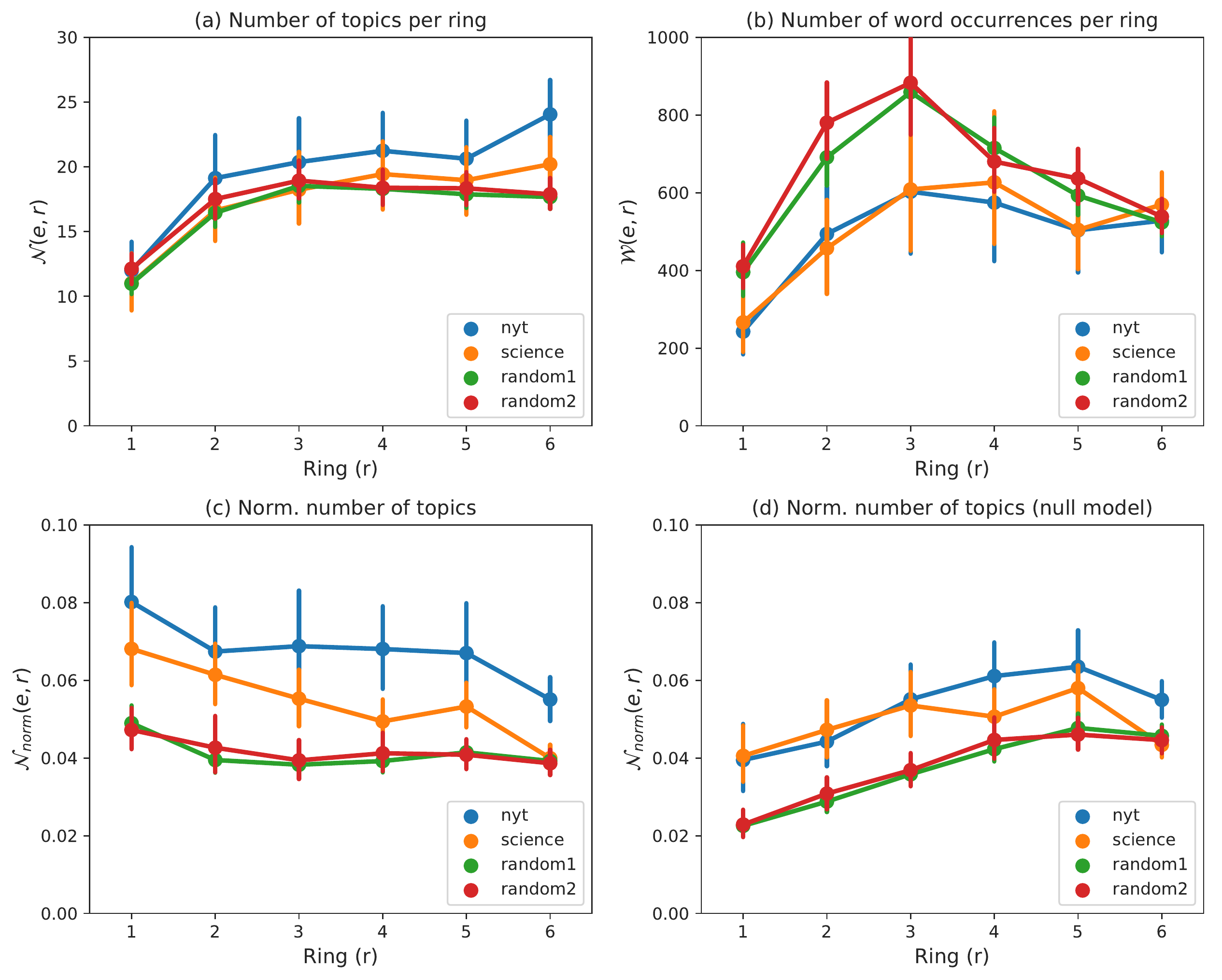}}
  \caption{Average number of topics (a), number of word occurrences (b), and normalised number of topics (c) in each ring of the ego network. For ``null'' ego networks, we report only the normalised number of topics (d).}
  \label{fig:topic-gen}
\end{figure}

In Fig.~\ref{fig:topic-gen} (a), we can observe that the number of topics grows towards the external rings (from about 11 in ring \#1 to over 16 in ring \#6). However, not all rings contain the same number of word occurrences (Fig.~\ref{fig:topic-gen} (b)): as seen previously in Section~\ref{sec:topic-extract}, each word occurrence contributes equally and independently to the calculation of the topics distribution. Therefore, a ring containing more word occurrences is more likely to contain more different topics. When we normalise by word occurrences ($\mathcal{N}_{norm}(r)$), the maximum of the normalised topic count (Fig.~\ref{fig:topic-gen} (c)) is observed in the first ring. Thus, \emph{ring \#1 stands out as the ring that generates proportionally more topics than the other rings}. 

In order to validate this hypothesis, we need to rule out that this result is not a mere side effect induced by the structure of the ego networks but it is a tell-tale sign of how humans pick the words in their innermost ring. In other words, we want to test whether keeping the ego network structure unchanged but swapping the words in the rings would still yield the same result regarding ring \#1. To this aim, we designed a null model where the ego network structure remains the same but the words are shuffled (more details in the grey box below).
In Fig.~\ref{fig:topic-gen} (d), we show $\mathcal{N}_{norm}(r)$ for the null model of ego networks. Since the maximum of $\mathcal{N}_{norm}(r)$ is obtained at a different ring $r$ than in the previous case, we can deduce that ring \#1 is special not just as a side effect of the ego network structure but due to the nature of the words it contains. To further confirm this finding, note also that the number of topics per word occurrence is significantly lower for innermost rings in the null model with respect to the outermost rings whereas the opposite is true for real ego networks. This is a second element that hints at the peculiar role of innermost rings in real-life ego networks of words.

\begin{figure}
\greybox{\textbf{Building a null model of an ego network.} \\
In order to show that the result is not only determined by the structure of the ego network (independently of the word organization inside), we chose to build ``null", artificial ego networks based on those already existing.
Let $o(w_u,e)$ be the number of occurrences of the word $w_u$ in ego $e$, such that the number of word occurrences in a ring $r$ of a given ego $e$ is defined as:

\begin{equation}
O(e,r) = \sum_{w_u \in W_u(e,r)} o(w_u,e),
\end{equation}

$\mathcal{W}_u(e,r)$ being the set of unique words in ring $r$. For each ego network, all the words are shuffled (i.e., a new $\mathcal{W'}_u$ is defined) and the word occurrences are artificially changed (new $o'$ and $O'$ are defined) such that the ring sizes and the number of occurrences are kept unchanged: 

\begin{equation}
    \label{eq:invariant-shuffle}
     \begin{cases}
        |\mathcal{W'}_u(e,r)| = |\mathcal{W}_u(e,r)| \\
        O'(e,r) = O(e,r).
    \end{cases}  
\end{equation}

The shuffling process can be considered as a succession of random swaps of words in the ego network. Let us consider a word $w_x$ with \textit{X} occurrences in ring $r_x$, and another word $w_x$  with \textit{Y} occurrences in ring $r_y$. During the shuffling process, assume the two words are swapped. In that new ego network, the number of occurrences of $w_x$ is forcibly set to the original number of occurrences of $w_y$ and vice versa:

\begin{equation}
    \begin{cases}
        o'(w_x,e) = o(w_y,e) = Y \\
        o'(w_y,e) = o(w_x,e) = X.
    \end{cases}
\end{equation}

That way, we can preserve Eq~\eqref{eq:invariant-shuffle}. Words are shuffled along with their topic distribution $P_{w_u}^{(e)}$ in the original dataset. This topic distribution associated to a unique word $w_u$ is calculated based on its occurrence $w \in \mathcal{W}(e,w_u)$. Each of these word occurrences $w$ is associated with a topic $c_w \in \mathcal{C}$ such that $P_{m(w_c)}(c)=1$. Hence, $P_{w_u}^{(e)}(c)$ simply corresponds to the ratio of the occurrences of $w_u$ that are associated to $c$.

\begin{equation}
    P_{w_u}^{(e)}(c) = \frac{1}{|\mathcal{W}(e,w_u)|} \sum_{w \in \mathcal{W}(e,w_u)} P_{m(w)}(c).
\end{equation}

Then the new topic distribution of a given ring $r$ is the weighted average of the topic distribution $P_{w_u}^{(e)}$ of the unique words $w_u \in \mathcal{W'}_u(e,r)$ that compose that ring after shuffling

\begin{equation}
    P_r^{(e)}(c) = \frac{\sum_{w_u \in \mathcal{W'}_u(e,r)} o'(w_u) \times P_{w_u}^{(e)}(c) }{ \sum_{w_u \in \mathcal{W'}_u(e,r)} o'(w_u) }.
\end{equation}

The full process is summarized with a toy example in Fig.~\ref{fig:null-model-shuffle}.



}
\end{figure}

\greybox{
\begin{figure}[H]
  \centering
  \iftoggle{NOFIG}{}{\includegraphics[width=\linewidth]{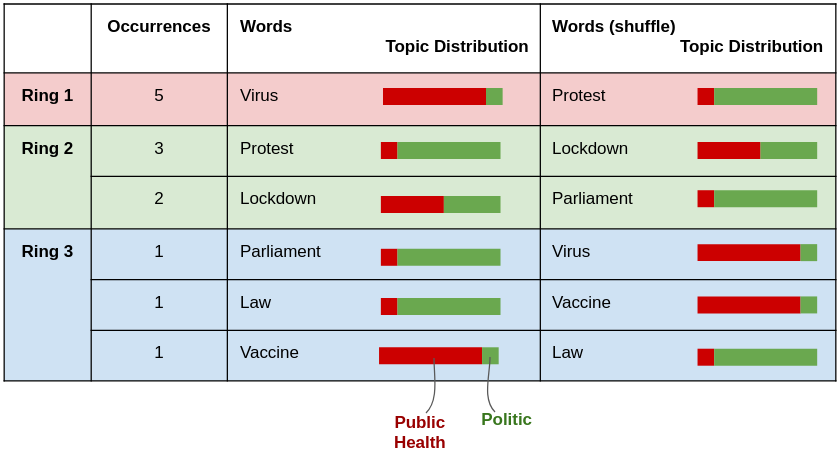}}
  \caption{{\bf Null model example}. The ring sizes and word occurrences are kept, the words are shuffled. In this toy example: $O(e,r_2)=3+2$, $o(virus,e)=5$, $o'(virus,e)=1$.}
  \label{fig:null-model-shuffle}
\end{figure}
}

\begin{figure}[H]
  \centering
  \iftoggle{NOFIG}{}{\includegraphics[width=\linewidth]{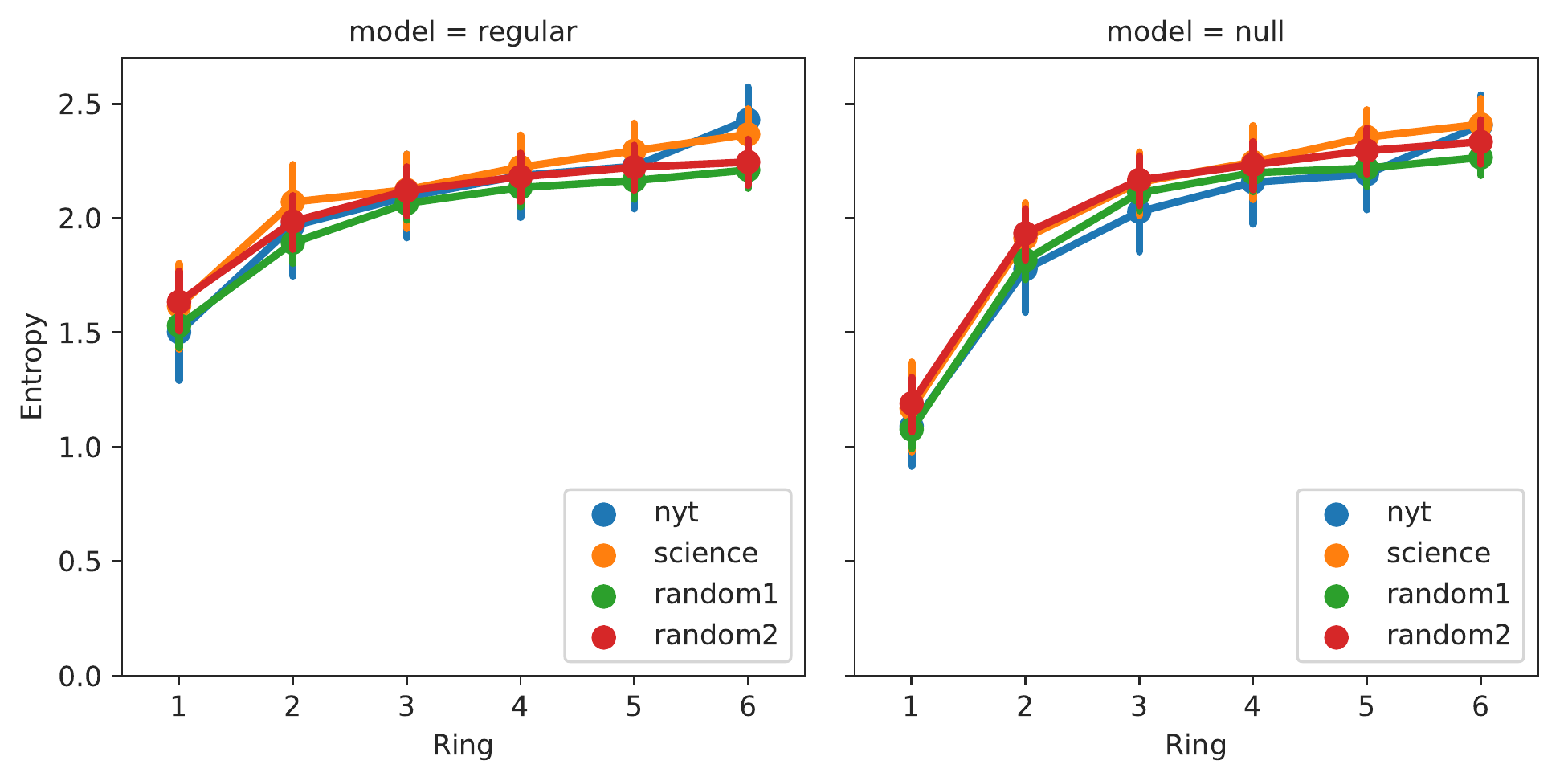}}
  \caption{{\bf Entropy of the semantic profiles per ring.}  Real-life ego networks (left) vs null model ego networks (right). \label{fig:entropy}}
\end{figure}

To extend our study beyond the mere number of topics per ring, we now investigate the diversity in the way topics are distributed, leveraging the entropy of the semantic profiles defined in Section~\ref{sec:sempro-charact}.
This is a way of calculating the semantic diversity of the words that compose a ring, as would be a metric like the average pairwise semantic distance, but based on the semantic profile that we have previously calculated.
Fig.~\ref{fig:entropy} (left) shows different levels of entropy depending on the rings: $H(r)$ grows towards the outer rings and is significantly lower in the innermost ring (for all datasets). This means that the outermost rings are, on average, semantically richer than the innermost ones. Then, we compare these results with those obtained from the null model (Fig.~\ref{fig:entropy} on the right), to find out whether the differences in entropy are related to the intrinsic structure of the ego network. We find that the entropy of the null model is the same as the original model for all rings, but for ring \#1, where the null model entropy is lower. \emph{This means that, even if words are organized in the ego network such that  the diversity of topics grows toward the outermost rings,  the diversity in ring \#1 is higher than what we could expect if words were randomly assigned to rings,} which is consistent with the previous findings of this section.

\begin{figure}[!h]
  \centering
  \iftoggle{NOFIG}{}{\includegraphics[width=\linewidth]{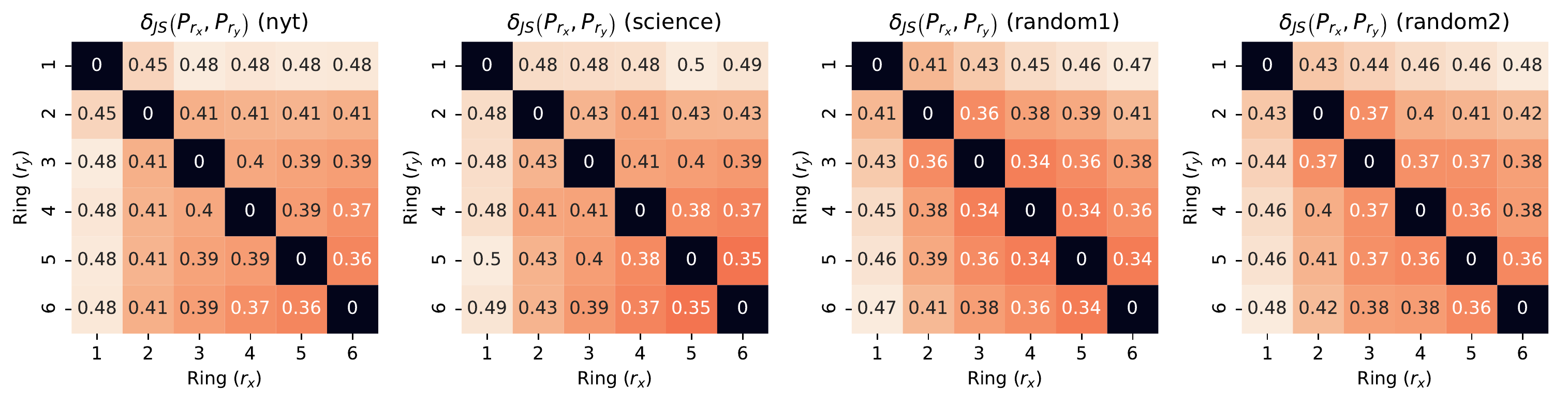}}
  \caption{{\bf Jensen-Shannon distance.} Average JS distance between the rings. \label{fig:jsd-table}}
\end{figure}

We now carry out a pairwise comparison of the semantic profiles of rings, using the JS distance described in Section~\ref{sec:sempro-ring-comparison}. we plot the, in Fig.~\ref{fig:jsd-table}. As one can expect, the diagonal is filled with zeros since the distance is calculated between two identical semantic profiles, and the upper triangle mirrors the lower triangle since the distance is symmetric. All datasets exhibit the same features: 
\begin{itemize}
\item The first row and column always contain the higher values. This means that ring \#1  (\textit{i.e.} the innermost ring) is always the most distant from the other rings. In other words, \emph{ring \#1 is the most characteristic ring.}
\item The lower values are always the distance between ring \#5 and \#6. Thus, \emph{the pairs of most similar rings are always among the outermost ones.}
\item For one row or column, the lowest value is always neighbouring the diagonal: given one ring $x$, the least distant ring is always the previous ring $x-1$ or the following one $x+1$. This means that \emph{two rings close to each other are more likely to be similar.}
\end{itemize}
The first observation is very important because it shows that the topic distribution associated with the most used words (those in the innermost ring) by a Twitter user is different from that associated with the least used words. This makes ring \#1 unique in two ways. \emph{It generates proportionally more topics than the others rings (Fig.~\ref{fig:topic-gen} (c)), but the distribution in ring \#1 is the furthest away from the others} (Fig.~\ref{fig:jsd-table}). This hints at a significantly higher ``semantic generative role" of inner rings as opposed to outer ones: each word occurring in an inner ring is able ``generate" more topics on which the user engages. And these topics, on which that user focuses most (inner rings feature higher frequency of use of words) generate a distribution that is quite distinct from the one at the outermost rings, on which the user engages far less.


\emph{Take home message for Section~\ref{sec:results-topics-per-ring}:} Ring \#1 is special in the ego network of words: it generates proportionally more topics than the other rings, its topic diversity is proportionally higher than expected, and its semantic profile is the most different with respect to the other rings. This suggests that ring \#1 may be the \emph{semantic fingerprint} of the ego network of words.

\subsubsection{The role of primary topics from ring \#1}
\label{sec:results-primary-topics-r1}

In the previous section, we discovered that ring \#1 is special. It, therefore, makes sense to investigate which topics are most important in this ring and if they tend to be equally important in the other rings. This will allow the reader to familiarize themselves with the methodology as well, before generalizing the analysis to other rings in Section~\ref{sec:results-pulling-power}.

\begin{figure}[!h]
  \centering
  \iftoggle{NOFIG}{}{\includegraphics[width=\linewidth]{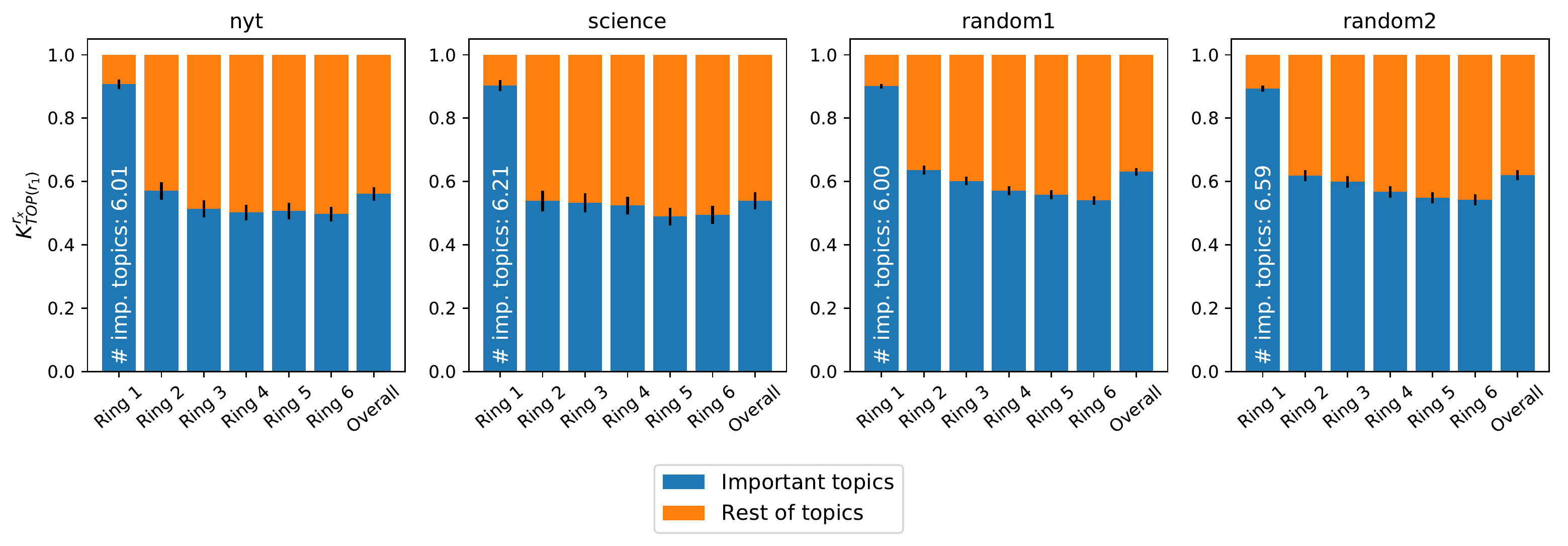}}
  \caption{{\bf Average strength of ring \#1's important topics in the semantic profile of each ring and of the whole ego network.} Each bar stands for the semantic profile of each ring (and overall ego network, in the last bar), where the blue part represents the share covered by the most important topics of ring \#1 (their average number $|U_{r_1}|$ is written in white).}
  \label{fig:r1-impt-topic-overall}
\end{figure}

We measure the overall importance of $r_1$'s primary topics in another ring $r_y$ by computing $K_{TOP(r_1)}^{r_y}$ (see Section~\ref{sec:important-topics}), varying $r_y$ from innermost to outermost layer. Fig.~\ref{fig:r1-impt-topic-overall} shows the coverage of $r_1$'s primary topics in the other rings, across all the ego networks. $K_{TOP(r_1)}^{r_y}$ corresponds to the blue bars in the figure. $K_{TOP(r_1)}^{r_y}$ accounts for approximately 50\% of each ring and of the whole ego network (last bar). This small (5-6, on average) set of topics, which fills almost the entire innermost ring, is playing a big role in the entire ego network as well. 

\begin{figure}[!h]
  \centering
  \iftoggle{NOFIG}{}{\includegraphics[width=\linewidth]{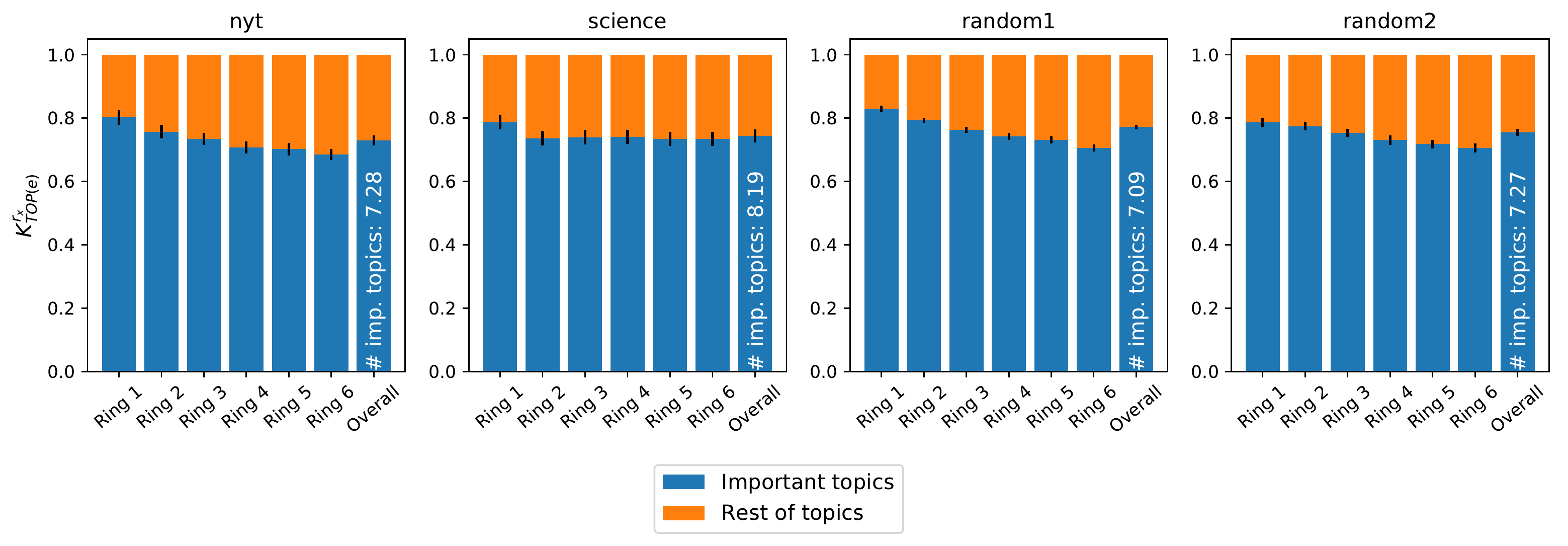}}
  \caption{{\bf Average strength of the ego network's important topics in the semantic profile of each ring.} The blue part of the stacked bar represents the share covered by the important topics in $U_e$. The average number of topics $|U_e|$ is specified in white.}
  \label{fig:overall-impt-topic-r1}
\end{figure}

To verify if the reverse statement is true (i.e., if topics that are important in the whole ego network are also important in ring \#1), we build a new set of topics $U_e$ grouping the most important topics in the whole ego network and calculate $K_{TOP(e)}^{r_y}$. Fig.~\ref{fig:overall-impt-topic-r1} highlights the coverage of those topics across the rings. Although, in general, all primary topics at the level of the ego network are well represented in all rings, we observe a slight predominance in ring \#1, as the innermost ring contains the biggest share of the most important topics of the ego network. This means that topics that are important to the ego network are over-represented in the innermost ring, i.e., an important topic discussed by a Twitter user is very likely to belong to $U_e^{r_1}$.

\emph{Take home message for Section~\ref{sec:results-primary-topics-r1}:}  Both results from Fig.~\ref{fig:r1-impt-topic-overall} and ~\ref{fig:overall-impt-topic-r1} indicate a close relation between important topics in ring \#1 and those important for the whole ego network. This observation is all the more interesting as ring \#1 is semantically the most different from all the others (Section~\ref{sec:results-topics-per-ring}), confirming the special role of this ring in the ego network of words.

\subsubsection{Pulling power of primary topics}
\label{sec:results-pulling-power}

Let us now focus on the primary topics in a generic ring $r_x$ (i.e., those in $U_{r_x}^{(e)}$). They can also appear in another ring $r_y$, and can be found in either $U_{r_y}^{(e)}$ or $L_{r_y}^{(e)}$. In the first case, the topics are primary in both rings, in the latter they are primary only in $r_x$.
We now tackle the following problem: which is the ring whose primary topics are most dominant among the primary topics of another ring? This involves measuring the strength, in the semantic profile of $r_y$, of the topics that are important for both $r_y$ and $r_x$. Using the notation of Section~\ref{sec:important-topics}, this is equivalent to studying $S_{TOP(r_x,r_y)}^{r_y}$ for all possible pairs of $r_x, r_y$. We show $S_{TOP(r_x,r_y)}^{r_y}$ on the left side of Table~\ref{tab:avg-distrib-per-ring}. The diagonal is left blank for the sake of clarity (we are interested in the results when $r_x \neq r_y$). For a given $r_y$, the largest value is written in bold.
We can clearly observe that the primary topics that are also primary in $r_1$ have almost always the largest share in the semantic profiles of the rings. Beyond the fact that the sum of important topics in ring \#1 is also important in the other rings (Section~\ref{sec:results-primary-topics-r1}), the table shows that they are on average the most likely to be important in all the other rings.

Now we tackle the complementary question: what is the pulling power of primary topics in a ring on the non-primary topics in another ring? We measure this via $S_{TOP(r_x),BOTTOM(r_y)}^{r_y}$, which is shown in the right part of  Table~\ref{tab:avg-distrib-per-ring}.

\begin{table}[!h]
\centering
 \caption{{\bf Pulling power of primary topics}. On the left, $S_{TOP(r_x,r_y)}^{r_y}$ for all $r_x,r_y$ pairs in our datasets. On the right, $S_{TOP(r_x),BOTTOM(r_y)}^{r_y}$. In bold, the highest value per column, corresponding to the $r_x$ for which the pulling power is higher in $r_y$.}
 \scriptsize
\setlength{\tabcolsep}{0.5em}
\renewcommand{\arraystretch}{1.2}
    \begin{tabular}{lccccccccccccc}
        \toprule
        & \multicolumn{13}{c}{Journalists} \\
        \cline{2-14}
        $r_x$ $r_y$ & \multicolumn{6}{c}{$S_{TOP(r_x,r_y)}^{r_y}$} & & \multicolumn{6}{c}{$S_{TOP(r_x),BOTTOM(r_y)}^{r_y}$} \\
        \cline{2-7} \cline{9-14}
        $\downarrow \rightarrow$ & $r_1$ & $r_2$ & $r_3$ & $r_4$ & $r_5$ & $r_6$ & & $r_1$ & $r_2$ & $r_3$ & $r_4$ & $r_5$ & $r_6$ \\ 
    
        $r_1$ &  & \textbf{.255} & \textbf{.226} & \textbf{.204} & .195 & \textbf{.180} & &  & .021 & .022 & \textbf{.023} & .022 & \textbf{.023} \\
        $r_2$ & .335 &  & .216 & .203 & .192 & .173 & & .025 &  & \textbf{.027} & \textbf{.023} & .030 & .022 \\
        $r_3$ & \textbf{.336} & .220 &  & .171 & \textbf{.196} & .162 & & \textbf{.026} & \textbf{.023} &  & .022 & \textbf{.032} & .020 \\
        $r_4$ & .321 & .230 & .190 &  & .167 & .154 & & .023 & .022 & \textbf{.027} &  & .029 & .022 \\
        $r_5$ & .307 & .235 & .209 & .184 &  & .151 & & .026 & \textbf{.023} & \textbf{.027} & \textbf{.023} &  & .022 \\
        $r_6$ & .318 & .234 & .210 & .188 & .179 &  & & .025 & .024 & \textbf{.027} & \textbf{.023} & .029 &  \\
        
        \cline{2-14}
        & \multicolumn{13}{c}{Science Writers} \\
        \cline{2-7} \cline{9-14}
        
        $r_1$ &  & .194 & \textbf{.191} & \textbf{.179} & \textbf{.169} & \textbf{.158} & &  & .023 & \textbf{.023} & .027 & .027 & .023 \\
        $r_2$ & .278 &  & .166 & .175 & .149 & .146 & & \textbf{.030} &  & .022 & .025 & .027 & \textbf{.024} \\
        $r_3$ & .285 & .172 &  & .154 & .153 & .146 & & .026 & \textbf{.026} &  & .024 & .028 & \textbf{.024} \\
        $r_4$ & .259 & \textbf{.200} & .169 &  & .147 & .148 & & .027 & .023 & .021 &  & .027 & \textbf{.024} \\
        $r_5$ & \textbf{.303} & .180 & .183 & .168 &  & .141 & & .027 & \textbf{.026} & .022 & \textbf{.028} &  & .023 \\
        $r_6$ & .253 & .193 & .183 & .171 & .150 &  & & .025 & .027 & .022 & .027 & \textbf{.029} &  \\
        
        \cline{2-14}
        & \multicolumn{13}{c}{Random Users \#1} \\
        \cline{2-7} \cline{9-14}
        
        $r_1$ &  & \textbf{.248} & \textbf{.216} & \textbf{.202} & \textbf{.203} & \textbf{.190} & &  & \textbf{.026} & .024 & .026 & .026 & .026 \\
        $r_2$ & \textbf{.284} &  & .202 & .192 & .189 & .178 & & \textbf{.030} &  & .025 & .027 & .026 & \textbf{.028} \\
        $r_3$ & .271 & .226 &  & .182 & .180 & .172 & & .028 & \textbf{.026} &  & \textbf{.028} & .026 & .027 \\
        $r_4$ & .259 & .214 & .188 &  & .177 & .168 & & .027 & .025 & \textbf{.026} &  & \textbf{.027} & .027 \\
        $r_5$ & .267 & .211 & .193 & .181 &  & .168 & & .028 & .025 & \textbf{.026} & .027 &  & .026 \\
        $r_6$ & .260 & .213 & .189 & .175 & .171 &  & & .028 & .023 & \textbf{.026} & .027 & .026 &  \\
        
        \cline{2-14}
        & \multicolumn{13}{c}{Random Users \#2} \\
        \cline{2-7} \cline{9-14}
        
        $r_1$ &  & \textbf{.222} & .199 & \textbf{.199} & \textbf{.179} & \textbf{.181} & &  & .024 & .021 & .025 & .020 & .025 \\
        $r_2$ & \textbf{.271} &  & \textbf{.203} & .187 & .177 & .178 & & .026 &  & .021 & .025 & .022 & .025 \\
        $r_3$ & .250 & .213 &  & .184 & .169 & .178 & & .025 & \textbf{.025} &  & \textbf{.026} & .021 & .025 \\
        $r_4$ & .255 & .202 & .191 &  & .168 & .165 & & \textbf{.027} & .024 & \textbf{.023} &  & \textbf{.023} & \textbf{.026} \\
        $r_5$ & .240 & .199 & .187 & .175 &  & .163 & & .025 & .023 & .022 & .025 &  & .025 \\
        $r_6$ & .246 & .207 & .190 & .178 & .158 &  & & .023 & .023 & .021 & .024 & .022 &  \\
        
    \end{tabular}
    \label{tab:avg-distrib-per-ring}
\end{table}

From the left side of Table~\ref{tab:avg-distrib-per-ring}, we know which is the ring whose primary topics have the highest pulling power on the primary topics of others. But do they have a higher than average strength with respect to the primary topics in the ring as a whole (i.e., regardless of whether they are primary in other rings or not)? To investigate this problem, we show $\sigma_{TOP(r_x,r_y)}^{r_y}$ in Table~\ref{tab:avg-distrib-per-ring-diff}. In the table, all the numbers are positive. This means that, on average, among the most important topics for a ring $r_y$, if a topic belongs to the important topics of another ring $r_x$, its strength will be more likely to be higher than the average strength of generic important topics in $r_y$. A $t$-test has been performed to assess whether these differences are statistically significant: in all cases, we obtained $p-\textrm{value}<.001$. On the right side of the table we show $\sigma_{TOP(r_x),BOTTOM(r_y)}^{r_y}$, which captures whether topics that are primary elsewhere but not in $r_y$ tend to have a higher share among the least important topics in $r_y$. In this case, too, the numbers are positive. It also means that, on average, among the least important topics of a given ring $r_y$, a topic is more likely to have a higher strength if it belongs to the important topics in another ring $r_x$. Again, the $p$-values are smaller than $.001$, confirming that such results are not due to statistical fluctuations.

\begin{table}[!h]
\centering
\caption{{\bf Pulling power of primary topics that are also primary elsewhere vs ``average'' primary / nonprimary topic.} On the left, $\sigma_{TOP(r_x,r_y)}^{r_y}$ for all $r_x,r_y$ pairs in our datasets. On the right, $\sigma_{TOP(r_x),BOTTOM(r_y)}^{r_y}$. The highest value per column is in bold.}
\scriptsize
\setlength{\tabcolsep}{0.5em}
\renewcommand{\arraystretch}{1.2}
    \begin{tabular}{lccccccccccccc}
        \toprule
        & \multicolumn{13}{c}{Journalists} \\
        \cline{2-14}
        $r_x$ $r_y$ & \multicolumn{6}{c}{$\sigma_{TOP(r_x,r_y)}^{r_y}$} & & \multicolumn{6}{c}{$\sigma_{TOP(r_x),BOTTOM(r_y)}^{r_y}$} \\
        \cline{2-7} \cline{9-14}
        $\downarrow \rightarrow$ & $r_1$ & $r_2$ & $r_3$ & $r_4$ & $r_5$ & $r_6$ & & $r_1$ & $r_2$ & $r_3$ & $r_4$ & $r_5$ & $r_6$ \\ 
        
        $r_1$ &  & \textbf{.059} & \textbf{.057} & \textbf{.068} & \textbf{.051} & \textbf{.058} & &  & \textbf{.006} & .007 & \textbf{.006} & .004 & \textbf{.005} \\
        $r_2$ & .082 &  & .044 & .060 & .043 & .051 & & \textbf{.006} &  & \textbf{.010} & .005 & .006 & .004 \\
        $r_3$ & \textbf{.090} & .035 &  & .040 & .036 & .039 & & .003 & \textbf{.006} &  & .004 & \textbf{.007} & .003 \\
        $r_4$ & .061 & .040 & .018 &  & .021 & .031 & & .003 & \textbf{.006} & .009 &  & .006 & .004 \\
        $r_5$ & .052 & .033 & .031 & .036 &  & .028 & & .005 & \textbf{.006} & \textbf{.010} & .004 &  & .003 \\
        $r_6$ & .061 & .032 & .027 & .029 & .018 &  & & .004 & .005 & .008 & .005 & .004 &  \\

        \cline{2-14}
        & \multicolumn{13}{c}{Science Writers} \\
        \cline{2-7} \cline{9-14}

        $r_1$ &  & .024 & \textbf{.048} & \textbf{.038} & \textbf{.043} & \textbf{.041} & &  & .002 & \textbf{.004} & \textbf{.006} & .004 & \textbf{.004} \\
        $r_2$ & \textbf{.035} &  & .033 & .027 & .022 & .025 & & \textbf{.004} &  & .003 & .005 & .003 & \textbf{.004} \\
        $r_3$ & .034 & \textbf{.025} &  & .019 & .027 & .026 & & .000 & .003 &  & .003 & .004 & .003 \\
        $r_4$ & .019 & \textbf{.025} & .034 &  & .019 & .027 & & .003 & .002 & .003 &  & .003 & \textbf{.004} \\
        $r_5$ & .045 & .022 & .037 & .020 &  & .021 & & .000 & .002 & .003 & .004 &  & .003 \\
        $r_6$ & .025 & .023 & .036 & .022 & .022 &  & & .002 & \textbf{.004} & \textbf{.004} & \textbf{.005} & .005 &  \\
        
        \cline{2-14}
        & \multicolumn{13}{c}{Random Users \#1} \\
        \cline{2-7} \cline{9-14}
        
        $r_1$ &  & \textbf{.063} & \textbf{.059} & \textbf{.049} & \textbf{.061} & \textbf{.053} & &  & \textbf{.006} & .004 & .006 & .004 & .002 \\
        $r_2$ & \textbf{.061} &  & .045 & .041 & .047 & .042 & & \textbf{.004} &  & .005 & .006 & .004 & \textbf{.004} \\
        $r_3$ & .045 & .039 &  & .032 & .037 & .036 & & \textbf{.004} & \textbf{.006} &  & \textbf{.007} & \textbf{.005} & \textbf{.004} \\
        $r_4$ & .035 & .033 & .032 &  & .034 & .031 & & .003 & .005 & \textbf{.006} &  & .004 & \textbf{.004} \\
        $r_5$ & .040 & .028 & .032 & .028 &  & .031 & & .003 & .005 & \textbf{.006} & .005 &  & \textbf{.004} \\
        $r_6$ & .035 & .032 & .033 & .023 & .028 &  & & \textbf{.004} & .004 & \textbf{.006} & .006 & .004 &  \\
        
        \cline{2-14}
        & \multicolumn{13}{c}{Random Users \#2} \\
        \cline{2-7} \cline{9-14}
        
        $r_1$ &  &\textbf{ .032} & \textbf{.043} & \textbf{.040} & \textbf{.048} & \textbf{.041} & &  & .005 & .005 & .004 & .002 & .003 \\
        $r_2$ & \textbf{.057} &  & .042 & .033 & \textbf{.048} & .038 & & .002 &  & .005 & .004 & \textbf{.003} & .002 \\
        $r_3$ & .041 & .024 &  & .029 & .037 & .037 & & .002 & \textbf{.006} &  & .004 & \textbf{.003} & .002 \\
        $r_4$ & .042 & .026 & .034 &  & .037 & .031 & & \textbf{.004} & .005 & \textbf{.006} &  & \textbf{.003} & \textbf{.004} \\
        $r_5$ & .029 & .019 & .025 & .020 &  & .023 & & .002 & .005 & .005 & \textbf{.005} &  & .002 \\
        $r_6$ & .031 & .022 & .029 & .024 & .026 &  & & .001 & .005 & .004 & .003 & .002 &  \\

    \end{tabular}
    \label{tab:avg-distrib-per-ring-diff}
\end{table}

\emph{Take home message for Section~\ref{sec:results-pulling-power}:}  Studying the role of primary topics, we have learned the following.
\begin{itemize}
    \item Primary topics from ring \#1 tend to dominate among the primary topics of other rings. This shows the pulling power of the innermost ring, confirming its special role in the ego network. Vice versa, primary topics from ring \#1 do not seem to dominate among non-primary topics of other rings.
    \item The topics that are primary in some rings tend to be stronger than average among the primary and non-primary topics in the semantic profile of another ring. This effect is especially acute when considering primary topics from ring \#1 with respect to generic primary topics in other rings.
\end{itemize}

\subsubsection{Discussion}
\label{sec:results-discussion}

The study of the semantic profile of the rings of the ego network confirms the relevance of the ego network of words model. This model allowed us to isolate the specific features of the topics associated with the words in the innermost ring. Indeed, the semantic profile in ring \#1 is not only the most unique (the most semantically distant from the others), but it is also characterized by both a larger than expected entropy distribution and number of topics generated, when compared with a null model.
The most important topics that ring \#1 is composed of are not only a set of important topics in the other rings: for every ring, an important topic is more likely to be predominant if it is also important in the innermost ring. Hence, despite the small number of unique words and word occurrences it contains, the innermost ring strongly ``predicts'' the most important topics in the entire ego network. \emph{In light of these results, we can conclude that the semantic profile of the innermost ring $r_1$ is also the semantic fingerprint of the whole ego network of words.}

As it has been done with social ego networks (using structural properties to study information diffusion~\cite{arnaboldi2017online}, or to perform link prediction~\cite{toprak2022harnessing}), we can use the structural and semantic invariants of the ego network of words to investigate some classical data science problems, with a focus on natural language processing. This semantic fingerprint could be used to identify specific Twitter users, or groups of users, with a non-trivial interest distribution for certain topics (e.g. a mix of important topics in the innermost rings and marginal topics in the outermost rings). It could also be used for link prediction with the assumption that users with the same topic of interest in the innermost ego network circles are more likely to follow one another (this is the principle of homophily) or for the purpose of word recommendation in a typing assistance tool. Since we identified some semantic invariants (eg. the role of important topics in ring \#1), we could leverage this property to identify outliers deviating from the standard and detect non-human behaviors. Finally, we could use the fact that ring \#1 contains the important topics of the entire ego network to spare some time considering only the words in this innermost ring, within the context of topic mining.

\section{Conclusion}
\label{sec:conclusion}

Inspired by previous work modeling the cognitive constraints that regulate personal social relations, in this paper, we investigate, through a data-driven approach, whether a regular structure can also be found in the way people use words, as a symptom of cognitive constraints in their mental process. 
%
Based on a corpus of tweets written by both regular and professional users, we have shown that, similarly to the social case, a concentric layered structure (which we name ``ego network of words'') very well captures how an individual organizes their cognitive effort in language production and reveals some structural invariants in the way people organise their own vocabulary. Among these invariants, we can list (i) the number of layers (between 5 and 7), (ii) their regular growth from the center of the word ego network outward (the innermost layer is five times smaller than the following one, for all the other layers their size approximately double moving outward), (iii) the size of external layers (which is pretty stable, with the two penultimate layers accounting respectively for 30\% and 60\% of the words in the model, regardless of the total number of layers). 

Then, going beyond words as units of language, we performed a semantic analysis of the ego network of words. Each ring of each ego network is described by a semantic profile that captures the topics associated with the words in the ring. We have found that ring \#1 has a special role in the model. It is semantically the most dissimilar out of the six, and also the one which generates proportionally the largest number of topics. We also showed that the topics that are important in the innermost ring, also have the characteristic of being predominant in each of the other rings, as well as in the entire ego network. In this respect, ring \#1 can be seen as the semantic fingerprint of the ego network of words. Finally, we found that the topics that are primary in some rings tend to be stronger than average among the primary and non-primary topics in the semantic profile of the other rings. This shows that, while layer \#1 provides a particularly strong signal about prevalence in the ego networks, weaker signals show a more complex structure of influence among topics ``resident" in different layers of the ego network of words.


 \subsubsection*{Acknowledgements.}

 This work was partially funded by the SoBigData++, HumaneAI-Net, and SAI projects. The SoBigData++ project has received funding from the European Union's Horizon 2020 research and innovation programme under grant agreement No 871042. The HumaneAI-Net project has received funding from the European Union's Horizon 2020 research and innovation programme under grant agreement No 952026.
 The SAI project is supported by the CHIST-ERA grant CHIST-ERA-19-XAI-010, by MUR (grant No. not yet available), FWF (grant No. I 5205), EPSRC (grant No. EP/V055712/1), NCN (grant No. 2020/02/Y/ST6/00064), ETAg (grant No. SLTAT21096), BNSF (grant No. KP-06-DOO2/5). 

 
\pagebreak




\appendix

\renewcommand{\thesection}{S1}

\renewcommand{\thesubsection}{S\arabic{section}.\arabic{subsection}}

\setcounter{table}{0}
\renewcommand{\thetable}{S\arabic{table}}

\captionsetup[figure]{labelfont={bf},name={Fig.},labelsep=period}
\setcounter{figure}{0}
\renewcommand{\thefigure}{S\arabic{figure}}

\section{Supporting information}
\label{S1-Appendix}

\subsection{Data preprocessing: filtering out inactive Twitter users}
\label{app:activeusers}

In order to be relevant to our work, a Twitter account must be an active account, which we define as an account not abandoned by its user and that tweets regularly. A Twitter account is considered abandoned, and we discard it, if the time since the last tweet is significantly bigger (we set this threshold at 6 months, as previously done also in~\cite{boldrini2018twitter}) than the largest period of inactivity for the account. We also consider the tweeting regularity, measured by counting the number of months where the user has been inactive. The account is tagged as sporadic, and discarded, if this number of months represents more than 50\% of the observation period (defined as the time between the first tweet of a user in our dataset and the download time). We also discard accounts whose entire timeline is covered by the 3200 tweets that we are able to download, because their Twitter behaviour might have yet to stabilise (it is known that the tweeting activity needs a few months after an account is created to stabilise).

\subsection{Ruling out soft clustering for the creation of semantic profiles}
\label{sec:soft-clustering-results}

\begin{figure}[H]
  \centering
  \iftoggle{NOFIG}{}{\includegraphics[width=\linewidth]{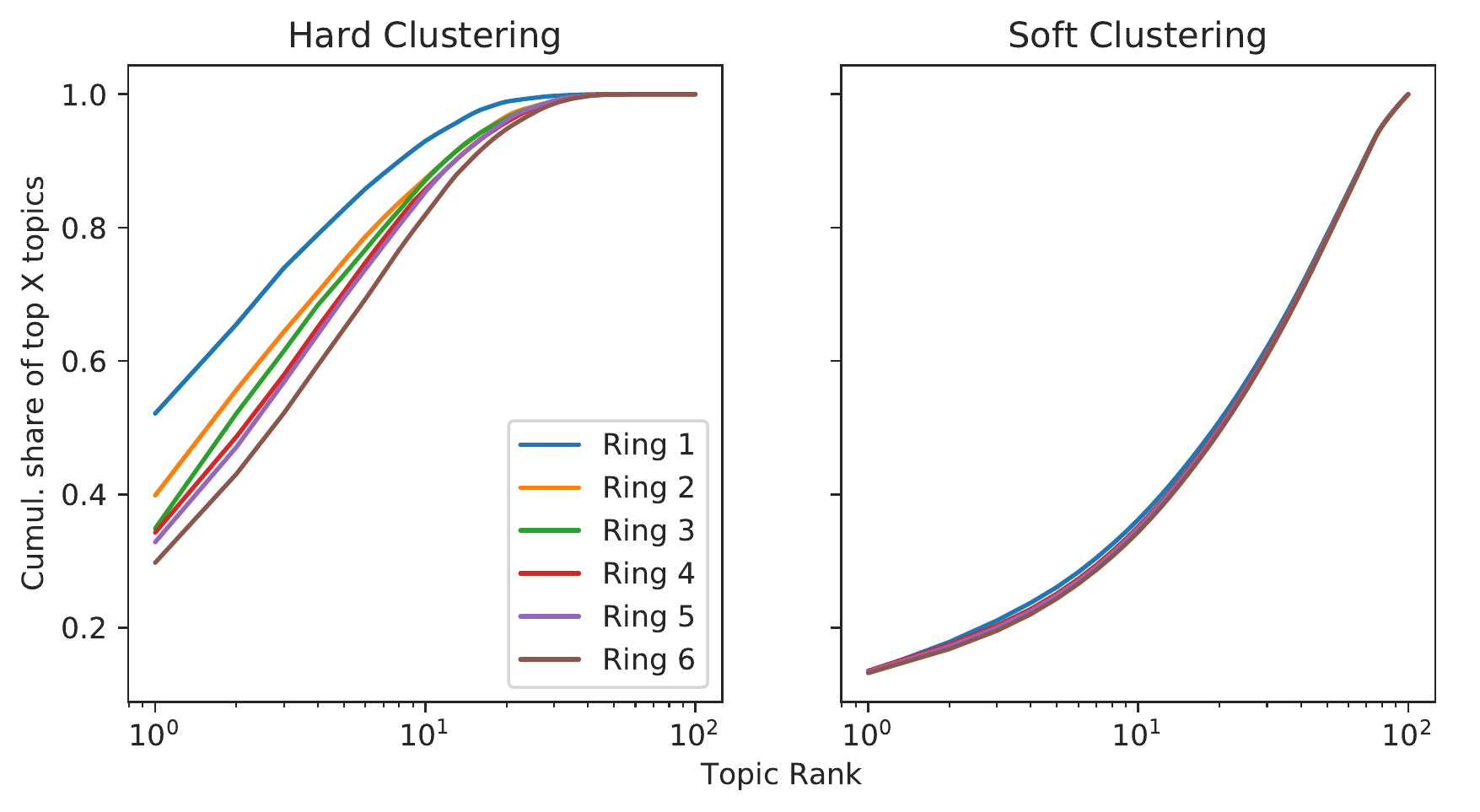}}
  \caption{{\bf Hard vs soft clustering.} Fraction of the semantic profile covered by the top-$x$ topics in the ring, after hard (left) and soft (right) clustering.}
  \label{fig:topics-cdf}
\end{figure}

In discussed in the body of the paper, the hard clustering approach to topic extraction yields many unassigned words (Table~\ref{tab:tweets-cluster}). We have thus also tested soft clustering, where by each word occurrence is assigned, in any case, a probability distribution of belonging to one of the 100 topics. In Fig~\ref{fig:topics-cdf} we plot the fraction of the semantic profile covered by the top-$x$ topics in the ring (where top-$x$ is computed based on the semantic profile $P_r^{(e)}$). 
Unlike hard clustering, soft clustering gives non-zero values to the least important topics of the ring. While soft clustering allows us to include all tweets in our analysis, it has a very negative side effect. As we show in the following of the section, very generic topics become prevalent, and mask more characteristic topics that hard clustering reveals, particularly for the innermost rings. Notice that this side effect makes all rings look alike in terms of number of active topics, as we can see from the fact that all distribution curves overlap in the right-hand side plots of Fig~\ref{fig:topics-cdf}.

\begin{figure}[H]
\iftoggle{NOFIG}{}{\includegraphics[width=\linewidth]{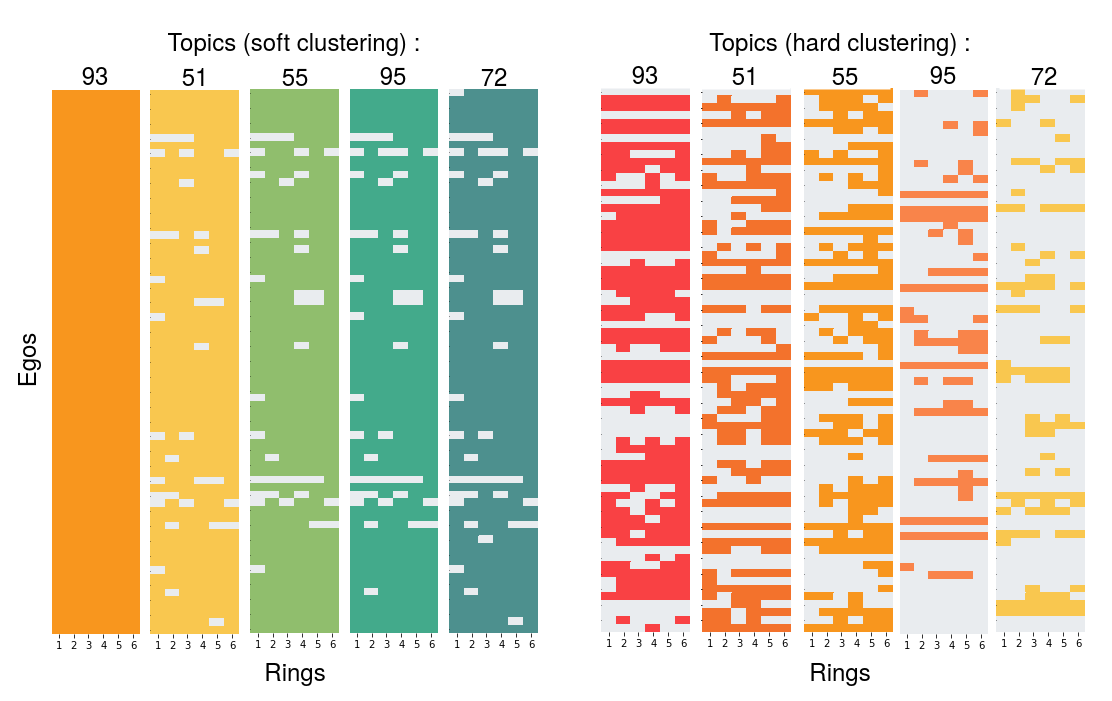}}
\caption{{\bf Hard vs soft clustering: five most dominant topics.} The two figures show how the five most important topics in the Journalists dataset are distributed, in the case of hard clustering (on the left) and soft clustering (on the right). For each topic, a grid is drawn in which the colored square means that the corresponding topic belongs to the most important topics of ring X of the ego network Y. Those topics are important for all six rings (the line is fully colored) for respectively 49\%, 28\%, 19\%, 21\%, 9\% of all the ego networks of the dataset for the hard clustered configuration (left) and 100\%, 75\%, 75\%, 74\%, 68\% for the soft clustered configuration.}
\label{fig:hardclust-sotclust}
\end{figure}

To better investigate this aspect, we extract the important topics as described in Section~\ref{sec:important-topics}. With two classes (important vs non-important), we obtain an average silhouette score of 0.9, confirming the good cluster configuration. We show these results for the Journalists dataset but similar conclusions can be drawn for the others. 
In Fig~\ref{fig:hardclust-sotclust}, we compare the level of importance of the 5 most dominant topics in the dataset (those who are important in the largest number of rings regardless of ego and ring rank), in the case of soft clustering and hard clustering.
The figure shows that soft clustering allows some topics to dominate the whole Journalists dataset. With soft clustering, topics 93, 51, 55, 95 and 72 are important for all six rings (the ego line is filled with colored squares) of more than 50\% of the ego networks. This, instead, is not the case when using hard clustering. The dominating topics in the case of soft clustering turn out being very generic ones. This is confirmed by looking at the most characteristic words in these topics in Table~\ref{tab:topic-impt-words}. For example topics 93 and 51, which were already among the most frequent in the hard cluster case are omnipresent in the soft cluster case, in addition to the topic 95 which is also generic but does not appear in the case of the hard cluster.
We can therefore conclude that the price of a complete inclusion of tweets in our topic analysis through soft clustering only increases the noise level for all ego networks, materialized by a set of very generic topics that blur the real semantic characteristics of the rings. This is why we decided to put aside the results related to the soft clustering, in order to keep only the semantic distributions resulting from the hard clustering of HDBSCAN. Note that, in light of these results, the fact that we use only a small subset of available tweets does not impact on the relevance of our analysis. What we exclude are the tweets related to ``noise" topics, in the sense that they are not able to strongly characterise the Twitter behaviour of users, and we focus only on tweets that are strongly belonging to topics, i.e., on the semantically characteristic part of users' Twitter activity.

\subsection{Additional tables}
\label{app:additionaltables}

\begin{table}[H]
\centering
\caption{{\bf Hashtags, links, emojis in the datasets.} In the process of word extraction, the tweet is decomposed in tokens which are usually separated by spaces. These tokens generally corresponds to words, but they can also be links, emojis and others markers that are specific to the online language such as hashtags. The table gives the percentage of hashtags, links and emojis, which are tokens filtered out from the datasets.}
\footnotesize
\setlength{\tabcolsep}{0.5em}
\renewcommand{\arraystretch}{1.2}
\begin{tabular}{lccc}
\toprule
                 & Percentage of hashtags & Percentage of links & Percentage of emojis \\
\midrule
Journalists      & 1.34 \%           & 7.27 \%        & 0.20 \%         \\
Science writers  & 3.47 \%           & 8.02 \%        & 0.55 \%         \\
Random users \#1 & 16.84 \%          & 6.97 \%        & 5.21 \%         \\
Random users \#2 & 7.20 \%           & 6.42 \%        & 4.60 \%        \\
\bottomrule
\end{tabular}
\label{tab:removedtokens}
\end{table}

\begin{table}[H]
\centering
\caption{{\bf Example of word extraction results.}}
    \scriptsize
    \setlength{\tabcolsep}{0.6em}
    \renewcommand{\arraystretch}{1.5}
    \begin{tabularx}{\textwidth}{XX}
        \toprule
        Original tweet content & List of words after pre-processing \\
        \midrule 
        The @Patriots say they don’t spy anymore. The @Eagles weren’t taking any chances. They ran a "fake" practice before the \#SuperBowl & spy, anymore, chance, run, fake, practice\\
        \#Paris attacks come 2 days before world leaders will meet in \#Turkey for the G20. Will be a huge test for Turkey. & attack, come, day, world, leader, meet, huge, test, turkey \\
        Latest garden species - the beautiful but destructive rosemary beetle, and a leafhopper (anyone know if this can be identified to species level from photo? Happy to give it a go)  \#30DaysWild \#MyWildCity \#gardening & late, garden, specie, beautiful, destructive, rosemary, beetle, leafhopper, know, identify, specie, level, photo, happy \\
        \bottomrule
    \end{tabularx}
    \label{tab:token-lemma}
\end{table}

\begin{table}[H]
\centering
\caption{{\bf Most characteristic words per topic.} They are obtained with a TF-IDF scoring.}
\scriptsize
\setlength{\tabcolsep}{0.5em}
\renewcommand{\arraystretch}{1.2}
    \begin{tabular}{ll}
        \toprule
            Topic & Characteristic words (TF-IDF) \\  \midrule
            64 & new obama administration tax white house comey donald president trump\\
            24 & cook lunch like dinner cheese chicken restaurant pizza food eat\\
            93 & boston read old like summer blue think google vega know\\
            62 & gop house obamacare vote repeal cut health senate republican tax\\
            51 & past february day tennis sentence week yesterday month ago year\\
			\midrule
           93  & boston read old like summer blue think google vega know\\
           51 & past february day tennis sentence week yesterday month ago year\\
           55  & london orleans nyc brooklyn statue monument time confederate new york\\
           95 & happy nice kind christmas great glad love thanks good thank\\
           72  & sharif judge state case pakistan gay execution supreme court arkansas\\
        \bottomrule
    \end{tabular}
   \label{tab:topic-impt-words}
\end{table}


\begin{table}[!p]
\centering
\caption{{\bf Topics of the NYT journalists dataset.} Most characteristic words and distribution in rings.}
\scriptsize
\setlength{\tabcolsep}{0.5em}
\renewcommand{\arraystretch}{1.2}
    \begin{tabular}{llcccccc}
        \toprule
Topic & Characteristic words (TF-IDF) & R1 & R2 & R3 & R4 & R5 & R6\\  \midrule
0 & australia australian story indigenous new & .006 & .008 & .009 & .005 & .005 & .003 \\
1 & yankee baseball game pitch hit & .010 & .007 & .011 & .011 & .012 & .011 \\
2 & italian soccer migrant libyan team & .009 & .010 & .004 & .009 & .004 & .006 \\
3 & alabama governor senate robert moore & .000 & .001 & .001 & .003 & .004 & .005 \\
4 & horse derby kentucky win race & .000 & .002 & .002 & .003 & .001 & .004 \\
5 & apple mac use new silver & .001 & .003 & .009 & .005 & .005 & .006 \\
6 & midwest south city today times & .013 & .010 & .010 & .012 & .014 & .010 \\
7 & fox news pope fake vatican & .015 & .009 & .010 & .006 & .006 & .008 \\
8 & french election macron pen paris & .000 & .000 & .008 & .004 & .002 & .001 \\
9 & white shark nationalist president harvard & .002 & .003 & .003 & .005 & .003 & .003 \\
10 & black slave african american asian & .014 & .024 & .021 & .021 & .022 & .018 \\
11 & turkey turkish referendum protester president & .001 & .003 & .004 & .001 & .002 & .001 \\
12 & cat mouse kitten game bureau & .001 & .004 & .006 & .002 & .004 & .005 \\
13 & birthday happy halloween spring valentine & .002 & .001 & .001 & .003 & .000 & .001 \\
14 & sleep bed nap asleep bedtime & .003 & .004 & .006 & .007 & .012 & .007 \\
15 & phone sorry storm stuck quick & .006 & .014 & .006 & .007 & .010 & .009 \\
16 & german right english angela fluent & .000 & .001 & .001 & .000 & .001 & .001 \\
17 & football bowl super player anthem & .004 & .007 & .008 & .005 & .006 & .007 \\
18 & brazil president brazilian scandal rio & .000 & .003 & .002 & .003 & .002 & .004 \\
19 & flight plane fly helicopter passenger & .002 & .002 & .002 & .001 & .002 & .002 \\
20 & beer vest clock declare power & .005 & .006 & .006 & .009 & .008 & .005 \\
21 & dog pet puppy love good & .004 & .005 & .006 & .006 & .005 & .005 \\
22 & wine red carpet school good & .007 & .008 & .008 & .007 & .006 & .004 \\
23 & fish boat surf fishing sea & .000 & .000 & .002 & .001 & .005 & .001 \\
24 & eat food pizza restaurant chicken & .001 & .003 & .004 & .005 & .008 & .006 \\
25 & train subway station new delay & .000 & .003 & .001 & .003 & .002 & .004 \\
26 & canada canadian refugee indigenous new & .001 & .001 & .002 & .001 & .001 & .002 \\
27 & year minute yahoo day hour & .014 & .013 & .011 & .009 & .006 & .005 \\
28 & bear montana wolf colorado wood & .003 & .006 & .003 & .004 & .004 & .004 \\
29 & hockey game team stanley cup & .000 & .002 & .001 & .000 & .001 & .000 \\
30 & snow ice winter cold arctic & .005 & .005 & .005 & .003 & .006 & .008 \\
31 & texas special state education cap & .008 & .004 & .005 & .007 & .006 & .006 \\
32 & sunday saturday night morning monday & .002 & .002 & .003 & .003 & .003 & .004 \\
33 & friday thursday tuesday monday wednesday & .002 & .003 & .016 & .006 & .006 & .007 \\
34 & moon space alien planet earth & .006 & .004 & .004 & .006 & .004 & .005 \\
35 & japan abe japanese reactor scandal & .003 & .005 & .013 & .008 & .010 & .011 \\
36 & china chinese hong new robot & .016 & .007 & .006 & .006 & .003 & .007 \\
37 & north missile korean nuclear south & .000 & .000 & .002 & .001 & .001 & .001 \\
38 & basketball league source trade season & .004 & .006 & .004 & .002 & .004 & .003 \\
39 & sigh mike right wow know & .005 & .009 & .005 & .004 & .003 & .003 \\
40 & twitter social medium like live & .003 & .002 & .003 & .002 & .002 & .003 \\
41 & miss destroyer sailor collision ship & .041 & .048 & .041 & .039 & .055 & .046 \\
42 & day july today year hour & .006 & .006 & .006 & .007 & .003 & .003 \\
43 & movie watch film play episode & .001 & .001 & .001 & .001 & .001 & .001 \\
44 & lobbyist intend dislike implication apology & .006 & .016 & .002 & .002 & .005 & .002 \\
45 & california earthquake san francisco quake & .009 & .003 & .002 & .001 & .000 & .000 \\
46 & hurricane florida irma storm harvey & .030 & .046 & .054 & .069 & .061 & .066 \\
47 & prince woman crown ebony ballroom & .015 & .007 & .010 & .001 & .006 & .004 \\
48 & iran iranian deal nuke president & .008 & .004 & .006 & .005 & .005 & .005 \\
49 & syrian attack chemical strike weapon & .001 & .003 & .013 & .008 & .005 & .010 \\
50 & russian russia trump investigation election & .001 & .000 & .002 & .000 & .002 & .001 \\
    \end{tabular}
   \label{tab:topics-nyt-1}
\end{table}

\begin{table}[!h]
\centering
\scriptsize
\setlength{\tabcolsep}{0.5em}
\renewcommand{\arraystretch}{1.2}
    \begin{tabular}{llcccccc}
51 & year ago month yesterday week & .009 & .016 & .013 & .017 & .010 & .015 \\
52 & climate trump change paris cut & .001 & .002 & .001 & .002 & .008 & .003 \\
53 & climate change oil paris carbon & .004 & .002 & .002 & .001 & .002 & .002 \\
54 & tweet chronological good evergreen great & .038 & .018 & .037 & .039 & .030 & .024 \\
55 & york new confederate time monument & .003 & .002 & .003 & .003 & .004 & .008 \\
56 & tax estate cash bank fund & .001 & .004 & .001 & .003 & .003 & .002 \\
57 & famine south yemen cholera venezuelan & .197 & .147 & .136 & .121 & .113 & .127 \\
58 & listen book talk daily new & .022 & .009 & .003 & .014 & .013 & .014 \\
59 & morning tomorrow good trial page & .007 & .008 & .006 & .009 & .011 & .008 \\
60 & week hour month year marathon & .010 & .008 & .010 & .007 & .008 & .010 \\
61 & million year billion spend marijuana & .004 & .007 & .006 & .007 & .007 & .005 \\
62 & tax republican senate health cut & .023 & .017 & .017 & .025 & .031 & .021 \\
63 & school high homework student college & .007 & .008 & .006 & .006 & .010 & .008 \\
64 & trump president donald house white & .003 & .005 & .003 & .002 & .002 & .003 \\
65 & big palestinian time read story & .005 & .002 & .003 & .006 & .002 & .005 \\
66 & fashion week mother wear model & .006 & .004 & .008 & .005 & .005 & .006 \\
67 & dress leather pink skirt gown & .022 & .022 & .025 & .026 & .023 & .029 \\
68 & tonight weekend atlanta bachelor georgia & .000 & .001 & .001 & .001 & .001 & .001 \\
69 & song hip rap rock hop & .005 & .001 & .001 & .000 & .000 & .001 \\
70 & broadway opera theater classical music & .006 & .012 & .002 & .002 & .002 & .003 \\
71 & dot reporter peer time & .026 & .017 & .011 & .021 & .016 & .017 \\
72 & arkansas court supreme execution gay & .006 & .012 & .010 & .012 & .009 & .010 \\
73 & sexual harassment woman accuse allegation & .024 & .034 & .039 & .024 & .016 & .025 \\
74 & wait bus happen mean depend & .038 & .010 & .007 & .003 & .003 & .002 \\
75 & send address question shoot reach & .000 & .000 & .000 & .001 & .001 & .001 \\
76 & car driver drive driving self & .001 & .003 & .001 & .001 & .002 & .001 \\
77 & eclipse solar total delete totality & .029 & .038 & .040 & .036 & .037 & .037 \\
78 & story news journalist accuse public & .070 & .071 & .084 & .085 & .115 & .104 \\
79 & suicide trial roy conrad carter & .001 & .003 & .005 & .005 & .005 & .006 \\
80 & die dead york robert roger & .001 & .001 & .005 & .008 & .004 & .003 \\
81 & roe squeamish lisa susan collins & .002 & .023 & .005 & .004 & .005 & .004 \\
82 & dislike unintended implication apology culture & .007 & .012 & .010 & .011 & .009 & .008 \\
83 & lady girl yes elizabeth finale & .003 & .004 & .010 & .008 & .007 & .007 \\
84 & book soon read write editor & .003 & .000 & .000 & .000 & .000 & .002 \\
85 & best great video love game & .006 & .012 & .008 & .010 & .010 & .012 \\
86 & bad terrible hate sorry awful & .002 & .005 & .008 & .005 & .005 & .006 \\
87 & drug police arrest jail gang & .020 & .016 & .016 & .014 & .020 & .020 \\
88 & kill militant police army congo & .000 & .003 & .007 & .005 & .003 & .005 \\
89 & agree tweet important fascinate interesting & .001 & .002 & .002 & .003 & .002 & .003 \\
90 & wrong bad argue moly mean & .003 & .005 & .011 & .011 & .007 & .010 \\
91 & love woman genius happy sandra & .004 & .005 & .005 & .005 & .002 & .004 \\
92 & yes true right joke correct & .002 & .001 & .002 & .004 & .004 & .003 \\
93 & know vega think blue summer & .011 & .012 & .011 & .011 & .015 & .016 \\
94 & beautiful great cool gorgeous fun & .010 & .012 & .005 & .004 & .008 & .006 \\
95 & good love glad great christmas & .024 & .030 & .015 & .023 & .019 & .019 \\
96 & god know exactly gold yes & .016 & .013 & .012 & .009 & .010 & .008 \\
97 & tho alex come like pat & .000 & .002 & .002 & .003 & .003 & .002 \\
98 & kate congratulation diane karen welcome & .018 & .012 & .014 & .028 & .016 & .017 \\
99 & read share contact matt paul & .003 & .003 & .003 & .006 & .003 & .002 \\
        \bottomrule
    \end{tabular}
   \label{tab:topics-nyt-2}
\end{table}


\begin{table}[!p]
\centering
\caption{{\bf Topics of the science writers dataset.} Most characteristic words and distribution in rings.}
\scriptsize
\setlength{\tabcolsep}{0.5em}
\renewcommand{\arraystretch}{1.2}
    \begin{tabular}{llcccccc}
        \toprule
Topic & Characteristic words (TF-IDF) & R1 & R2 & R3 & R4 & R5 & R6\\  \midrule
0 & daily late luck today & .002 & .002 & .002 & .003 & .000 & .003 \\
1 & baseball game lacrosse football player & .008 & .008 & .006 & .007 & .009 & .011 \\
2 & follower week new canada right & .014 & .013 & .011 & .017 & .013 & .016 \\
3 & video subtitle individual anonymous credit & .009 & .005 & .006 & .002 & .007 & .005 \\
4 & aku morning good river countryside & .000 & .000 & .000 & .000 & .000 & .000 \\
5 & aku morning good lake photo & .010 & .004 & .002 & .004 & .005 & .002 \\
6 & badge earn level middle road & .009 & .022 & .017 & .009 & .006 & .005 \\
7 & web nature post life plastic & .000 & .000 & .000 & .000 & .000 & .000 \\
8 & essay environmental educator nature conservation & .009 & .008 & .004 & .006 & .007 & .006 \\
9 & daily late clow soon hourly & .003 & .003 & .002 & .005 & .003 & .003 \\
10 & submission album cheer shoot hello & .018 & .008 & .014 & .012 & .019 & .023 \\
11 & submission album cheer shoot hello & .003 & .008 & .011 & .007 & .008 & .008 \\
12 & poker play chess player best & .011 & .006 & .006 & .006 & .005 & .006 \\
13 & robot human killer new job & .010 & .008 & .008 & .005 & .006 & .007 \\
14 & year gorilla monkey story ape & .003 & .008 & .008 & .007 & .006 & .007 \\
15 & white male quote diversity cause & .004 & .010 & .010 & .016 & .015 & .016 \\
16 & christmas holiday year tree festive & .034 & .049 & .033 & .045 & .044 & .056 \\
17 & plane flight fly spy airplane & .001 & .005 & .003 & .005 & .003 & .004 \\
18 & eclipse space moon earth solar & .004 & .002 & .003 & .002 & .003 & .003 \\
19 & african ancient beard genome revisit & .002 & .003 & .004 & .008 & .008 & .005 \\
20 & air asthma pollution risk city & .003 & .002 & .002 & .002 & .002 & .002 \\
21 & coffee shop drink caffeine cup & .024 & .029 & .022 & .047 & .029 & .031 \\
22 & drink beer brewery beach ale & .001 & .000 & .000 & .000 & .000 & .000 \\
23 & china chinese european scientific british & .003 & .006 & .003 & .003 & .002 & .002 \\
24 & morning good perambulation wake bob & .003 & .007 & .009 & .006 & .011 & .007 \\
25 & week virology new wildlife picture & .000 & .000 & .001 & .000 & .000 & .001 \\
26 & negotiation britain tax british european & .013 & .017 & .007 & .009 & .005 & .005 \\
27 & car driving self auto test & .046 & .038 & .038 & .025 & .027 & .026 \\
28 & happy birthday year mother wedding & .004 & .004 & .008 & .005 & .004 & .008 \\
29 & twitter mention reach social medium & .003 & .004 & .003 & .003 & .004 & .003 \\
30 & apple mobile search phone new & .008 & .005 & .006 & .007 & .005 & .009 \\
31 & weekly microbiology science episode new & .000 & .001 & .000 & .001 & .000 & .000 \\
32 & social medium fake news combat & .011 & .012 & .011 & .008 & .009 & .007 \\
33 & record hot year high warm & .039 & .029 & .043 & .056 & .050 & .056 \\
34 & journalist join hear sally tonight & .007 & .010 & .007 & .012 & .009 & .011 \\
35 & prize chemistry win medicine physiology & .006 & .005 & .008 & .005 & .005 & .006 \\
36 & chicken meat eat animal barn & .000 & .004 & .003 & .007 & .004 & .004 \\
37 & sleep bed night nap dream & .034 & .036 & .013 & .015 & .015 & .018 \\
38 & earthquake quake tsunami seismic big & .003 & .006 & .009 & .005 & .005 & .005 \\
39 & ice arctic winter snow antarctica & .005 & .011 & .009 & .007 & .007 & .008 \\
40 & canada canadian maple citizenship government & .005 & .005 & .012 & .011 & .006 & .009 \\
41 & california wildfire northern burn flee & .004 & .013 & .006 & .005 & .009 & .008 \\
42 & hurricane storm flood rain irma & .006 & .008 & .012 & .009 & .009 & .010 \\
43 & old fossil human year ancient & .001 & .004 & .005 & .004 & .007 & .005 \\
44 & frog otter snake amphibian rid & .069 & .067 & .071 & .067 & .076 & .075 \\
45 & pterosaur skull crest cornified animal & .005 & .002 & .003 & .004 & .003 & .001 \\
46 & bird spider bat flower moth & .004 & .009 & .014 & .011 & .012 & .015 \\
47 & dinosaur fossil bird mammal discover & .119 & .107 & .146 & .137 & .148 & .142 \\
48 & shark whale sea fish ocean & .002 & .002 & .005 & .004 & .004 & .006 \\
49 & bear wolf polar kill rhino & .006 & .007 & .008 & .008 & .010 & .011 \\
50 & dog puppy good breed love & .002 & .002 & .003 & .002 & .001 & .001 \\
    \end{tabular}
   \label{tab:topics-science-1}
\end{table}

\begin{table}[!h]
\centering
\scriptsize
\setlength{\tabcolsep}{0.5em}
\renewcommand{\arraystretch}{1.2}
    \begin{tabular}{llcccccc}
51 & chocolate eat pizza pie cheese & .002 & .003 & .003 & .002 & .002 & .003 \\
52 & food delicious fortune restaurant love & .009 & .002 & .003 & .003 & .002 & .002 \\
53 & cat dog kitten like think & .037 & .039 & .032 & .032 & .035 & .035 \\
54 & rule tobacco regulatory million health & .000 & .000 & .000 & .000 & .000 & .000 \\
55 & year time hour paper china & .008 & .007 & .007 & .007 & .004 & .002 \\
56 & woman award stem girl winner & .001 & .002 & .003 & .003 & .002 & .004 \\
57 & editor story wired write business & .041 & .033 & .043 & .044 & .038 & .041 \\
58 & car bicycle bike crash driving & .001 & .002 & .002 & .001 & .004 & .002 \\
59 & solar power wind energy electricity & .000 & .000 & .000 & .000 & .000 & .000 \\
60 & american america black prescription slavery & .001 & .001 & .005 & .003 & .004 & .006 \\
61 & die child woman bad parent & .020 & .011 & .008 & .009 & .013 & .011 \\
62 & health medical care patient doctor & .009 & .005 & .014 & .007 & .007 & .010 \\
63 & photo pic sharpen color apply & .004 & .005 & .007 & .007 & .007 & .008 \\
64 & cancer new cell mouse disease & .004 & .002 & .001 & .003 & .008 & .005 \\
65 & chromosome human horse embryo gene & .003 & .005 & .006 & .004 & .003 & .003 \\
66 & republican senate house senator white & .017 & .008 & .008 & .009 & .007 & .006 \\
67 & year day week halloween time & .002 & .000 & .000 & .002 & .001 & .000 \\
68 & trump administration president climate donald & .002 & .005 & .002 & .004 & .003 & .003 \\
69 & nuclear north weapon war iran & .003 & .006 & .001 & .003 & .003 & .002 \\
70 & coal oil climate fuel kentucky & .004 & .004 & .012 & .005 & .012 & .006 \\
71 & climate change carbon scientist report & .007 & .003 & .003 & .003 & .004 & .005 \\
72 & defense arrive plant episode week & .001 & .008 & .009 & .005 & .007 & .006 \\
73 & kill police murder arrest officer & .010 & .009 & .011 & .012 & .009 & .007 \\
74 & documentary film watch new series & .013 & .014 & .003 & .005 & .002 & .001 \\
75 & year end hour ago chronicle & .000 & .000 & .000 & .000 & .000 & .000 \\
76 & send address dot touch chat & .004 & .002 & .001 & .003 & .002 & .004 \\
77 & science donation match great recur & .009 & .003 & .004 & .005 & .006 & .008 \\
78 & like good way think know & .001 & .004 & .002 & .003 & .003 & .001 \\
79 & boston stereo arena queen wed & .021 & .016 & .008 & .010 & .009 & .008 \\
80 & great year sing night happy & .000 & .001 & .007 & .003 & .006 & .001 \\
81 & night stream miss catch tonight & .005 & .004 & .006 & .005 & .005 & .004 \\
82 & week month year new tomorrow & .001 & .005 & .005 & .004 & .004 & .003 \\
83 & science student school week scientist & .006 & .020 & .011 & .013 & .019 & .010 \\
84 & sunday saturday night come need & .053 & .041 & .044 & .032 & .037 & .029 \\
85 & thursday friday join wednesday tuesday & .001 & .001 & .002 & .002 & .002 & .002 \\
86 & science sexual harassment obituary journalism & .011 & .011 & .005 & .008 & .007 & .006 \\
87 & community follow rank step work & .010 & .007 & .012 & .010 & .009 & .012 \\
88 & free article site tweet want & .011 & .009 & .008 & .006 & .009 & .006 \\
89 & book read weekend science journal & .000 & .000 & .000 & .000 & .000 & .000 \\
90 & mean think worry thing point & .000 & .004 & .001 & .002 & .004 & .004 \\
91 & know right sure check want & .012 & .020 & .008 & .010 & .006 & .009 \\
92 & bad people medium crazy like & .003 & .001 & .001 & .001 & .001 & .001 \\
93 & sorry bad terrible sad weird & .038 & .023 & .023 & .031 & .021 & .020 \\
94 & god nope test idea know & .015 & .014 & .011 & .009 & .012 & .009 \\
95 & yes agree wow mean whoa & .005 & .007 & .008 & .012 & .009 & .011 \\
96 & glad kind great love enjoy & .000 & .004 & .003 & .006 & .001 & .001 \\
97 & good awesome love cool nice & .002 & .001 & .003 & .002 & .003 & .002 \\
98 & fan week big congratulation mull & .000 & .000 & .000 & .000 & .000 & .000 \\
99 & bless andy congratulation paul mate & .003 & .003 & .004 & .003 & .003 & .002 \\
        \bottomrule
    \end{tabular}
   \label{tab:topics-science-2}
\end{table}


\begin{table}[!h]
\centering
\caption{{\bf Topics of the random users \#1 dataset.} Most characteristic words and distribution in rings.}
\scriptsize
\setlength{\tabcolsep}{0.5em}
\renewcommand{\arraystretch}{1.2}
    \begin{tabular}{llcccccc}
        \toprule
Topic & Characteristic words (TF-IDF) & R1 & R2 & R3 & R4 & R5 & R6\\  \midrule
0 & twitter mention reach week like & .003 & .002 & .002 & .002 & .003 & .002 \\
1 & automatically unfollowed check follow people & .005 & .004 & .008 & .004 & .005 & .004 \\
2 & natural naturally soon tune launch & .004 & .005 & .005 & .004 & .005 & .005 \\
3 & post photo atlantic raw valley & .011 & .012 & .012 & .015 & .017 & .017 \\
4 & week fan big boy great & .001 & .001 & .001 & .001 & .001 & .001 \\
5 & replacement screen ram core battery & .005 & .004 & .005 & .005 & .005 & .006 \\
6 & practice spanish read news post & .007 & .007 & .007 & .007 & .008 & .008 \\
7 & bristol story chronicle daily include & .001 & .001 & .001 & .001 & .001 & .001 \\
8 & cannabis marijuana medical weed industry & .002 & .002 & .002 & .003 & .004 & .003 \\
9 & australia visa immigration australian apply & .006 & .002 & .002 & .001 & .001 & .001 \\
10 & alert trance dance hit triple & .007 & .007 & .007 & .005 & .005 & .005 \\
11 & hire job post pro apply & .000 & .002 & .001 & .002 & .004 & .004 \\
12 & music available game prophesy gospel & .000 & .002 & .002 & .002 & .002 & .002 \\
13 & happy peep good thanksgiving holiday & .022 & .025 & .029 & .028 & .026 & .025 \\
14 & canada immigration apply express entry & .011 & .013 & .006 & .004 & .005 & .006 \\
15 & visit information weekly clue chat & .001 & .003 & .001 & .001 & .001 & .001 \\
16 & track rock follower today outlaw & .001 & .002 & .002 & .003 & .004 & .004 \\
17 & late daily innovative horse source & .005 & .004 & .004 & .005 & .004 & .004 \\
18 & moon space mar astronaut mission & .009 & .004 & .004 & .002 & .002 & .001 \\
19 & road gold world win champ & .021 & .017 & .013 & .012 & .008 & .009 \\
20 & link subscribe click channel registration & .003 & .005 & .005 & .005 & .004 & .005 \\
21 & red blue sugar mug titan & .003 & .003 & .004 & .004 & .005 & .004 \\
22 & catholic priest pope church prayer & .000 & .000 & .001 & .001 & .001 & .001 \\
23 & trading risky suitable net close & .003 & .002 & .003 & .003 & .003 & .005 \\
24 & life sunday breath coach insurance & .004 & .007 & .005 & .005 & .005 & .004 \\
25 & god lord jesus christ unto & .005 & .003 & .003 & .003 & .002 & .002 \\
26 & associate page log principal excerpt & .001 & .002 & .003 & .002 & .002 & .003 \\
27 & amazon offer bank discount author & .029 & .020 & .017 & .012 & .014 & .011 \\
28 & phone car tune today mobile & .006 & .003 & .007 & .008 & .009 & .011 \\
29 & car hire plate vat drive & .006 & .008 & .011 & .007 & .008 & .009 \\
30 & christmas merry gift festive day & .001 & .003 & .003 & .003 & .005 & .005 \\
31 & friday weekend happy day halloween & .001 & .001 & .001 & .001 & .002 & .002 \\
32 & tea beer drink come brewery & .001 & .001 & .001 & .001 & .002 & .001 \\
33 & yoga teacher japanese meditation training & .006 & .006 & .007 & .006 & .007 & .006 \\
34 & life weight lose people think & .017 & .016 & .016 & .018 & .018 & .019 \\
35 & black white american fear legging & .008 & .010 & .009 & .008 & .009 & .008 \\
36 & today evangelist shower angela help & .009 & .009 & .012 & .013 & .013 & .014 \\
37 & thing dream life right time & .020 & .028 & .026 & .027 & .029 & .029 \\
38 & monday week morning happy good & .004 & .006 & .009 & .007 & .007 & .007 \\
39 & coffee cup morning good day & .009 & .017 & .009 & .007 & .005 & .005 \\
40 & password best wednesday frustration day & .004 & .003 & .004 & .004 & .004 & .003 \\
41 & dog pet puppy love dane & .006 & .003 & .004 & .004 & .005 & .004 \\
42 & cat kitten home lover happy & .009 & .008 & .011 & .013 & .013 & .014 \\
43 & apply badge level earn job & .019 & .006 & .006 & .004 & .003 & .003 \\
44 & tuesday today day good life & .004 & .006 & .005 & .006 & .005 & .005 \\
45 & food breakfast eat recipe chris & .013 & .017 & .015 & .019 & .022 & .021 \\
46 & cake chocolate cream ice birthday & .006 & .006 & .009 & .006 & .006 & .008 \\
47 & look nice delicious yummy forward & .006 & .005 & .008 & .008 & .009 & .011 \\
48 & flight dana fly update gate & .009 & .008 & .009 & .010 & .008 & .010 \\
49 & chicken curry lunch green menu & .005 & .005 & .003 & .004 & .003 & .004 \\
50 & follow hey kindly smile fib & .010 & .009 & .011 & .013 & .011 & .010 \\
    \end{tabular}
   \label{tab:topics-random1-1}
\end{table}

\begin{table}[!h]
\centering
\scriptsize
\setlength{\tabcolsep}{0.5em}
\renewcommand{\arraystretch}{1.2}
    \begin{tabular}{llcccccc}
51 & bedroom home house pool village & .031 & .032 & .036 & .035 & .034 & .034 \\
52 & shop fashion dress wedding buy & .008 & .008 & .006 & .005 & .007 & .005 \\
53 & cricket win match wicket cup & .005 & .004 & .004 & .005 & .004 & .004 \\
54 & win rocket game final score & .001 & .001 & .001 & .001 & .001 & .001 \\
55 & basketball football team game soccer & .055 & .051 & .056 & .055 & .053 & .057 \\
56 & beautiful cute hope bird look & .004 & .001 & .001 & .001 & .001 & .001 \\
57 & sorry inconvenience contact hear team & .006 & .006 & .008 & .008 & .006 & .007 \\
58 & sleep bed night wake nap & .005 & .006 & .006 & .006 & .007 & .005 \\
59 & winter snow cold ski rain & .002 & .001 & .002 & .003 & .002 & .002 \\
60 & tonight winner night win ticket & .013 & .011 & .008 & .009 & .006 & .006 \\
61 & connect let follow group family & .008 & .009 & .009 & .010 & .011 & .011 \\
62 & social medium hilarious engagement marketing & .010 & .004 & .005 & .003 & .002 & .002 \\
63 & live music official video bad & .022 & .024 & .025 & .027 & .029 & .025 \\
64 & dance befit class studio join & .001 & .001 & .001 & .002 & .002 & .002 \\
65 & video learn color alphabet child & .006 & .006 & .006 & .006 & .005 & .006 \\
66 & content write writer currently start & .011 & .010 & .009 & .009 & .009 & .007 \\
67 & climate east change late south & .057 & .038 & .036 & .033 & .033 & .035 \\
68 & stay park hostel hotel board & .015 & .022 & .019 & .024 & .020 & .019 \\
69 & oil climate join fossil fuel & .089 & .103 & .104 & .106 & .098 & .093 \\
70 & birthday happy wish bless year & .009 & .006 & .009 & .006 & .007 & .006 \\
71 & morning good golf bless day & .008 & .009 & .008 & .008 & .010 & .008 \\
72 & address send hello look certainly & .001 & .001 & .001 & .001 & .001 & .002 \\
73 & staff dudley health nurse mental & .008 & .010 & .011 & .011 & .010 & .011 \\
74 & vulnerable rat outstanding agency child & .002 & .002 & .004 & .005 & .007 & .006 \\
75 & help miss autism interested locate & .004 & .006 & .003 & .003 & .003 & .003 \\
76 & tutor tip directory literacy foot & .003 & .001 & .001 & .002 & .001 & .001 \\
77 & cancer patient therapy cell treatment & .006 & .003 & .000 & .000 & .000 & .000 \\
78 & west movie blast film watch & .023 & .022 & .026 & .026 & .026 & .024 \\
79 & million year store billion investment & .001 & .002 & .001 & .001 & .001 & .001 \\
80 & new salary happy year profile & .002 & .001 & .002 & .003 & .002 & .002 \\
81 & appreciate share shout homeless tweet & .003 & .002 & .002 & .002 & .002 & .002 \\
82 & school exam dismissal free generate & .016 & .021 & .019 & .019 & .018 & .017 \\
83 & mother brother son queen love & .011 & .012 & .013 & .015 & .016 & .017 \\
84 & book savvy silly society update & .004 & .004 & .004 & .004 & .005 & .005 \\
85 & woman ass sexy sensual sophisticated & .015 & .018 & .018 & .018 & .014 & .015 \\
86 & day verse valentine great grateful & .002 & .002 & .003 & .003 & .001 & .002 \\
87 & cloud marketing digital network robot & .106 & .105 & .104 & .107 & .111 & .111 \\
88 & career business support information employer & .001 & .003 & .002 & .002 & .001 & .002 \\
89 & year wait month code week & .006 & .009 & .007 & .005 & .005 & .004 \\
90 & welcome sacrifice salute nancy champagne & .009 & .010 & .014 & .015 & .015 & .016 \\
91 & love congratulation feedback great hug & .003 & .005 & .005 & .004 & .004 & .004 \\
92 & creation awesome create think look & .018 & .019 & .020 & .021 & .024 & .025 \\
93 & india anniversary indian birth kashmiri & .017 & .013 & .009 & .008 & .007 & .008 \\
94 & amen preach & .004 & .004 & .005 & .005 & .006 & .006 \\
95 & dream true agree old believe & .003 & .003 & .003 & .005 & .005 & .004 \\
96 & vote know people hold yes & .006 & .009 & .008 & .008 & .011 & .012 \\
97 & trump president russia hillary lawyer & .007 & .004 & .007 & .006 & .006 & .005 \\
98 & arrest police man kill murder & .000 & .001 & .001 & .002 & .001 & .002 \\
99 & bad sad disrespectful awful disgust & .000 & .000 & .000 & .000 & .000 & .000 \\
        \bottomrule
    \end{tabular}
   \label{tab:topics-random1-2}
\end{table}


\begin{table}[!h]
\centering
\caption{{\bf Topics of the random users \#2 dataset.} Most characteristic words and distribution in rings.}
\scriptsize
\setlength{\tabcolsep}{0.5em}
\renewcommand{\arraystretch}{1.2}
    \begin{tabular}{llcccccc}
        \toprule
Topic & Characteristic words (TF-IDF) & R1 & R2 & R3 & R4 & R5 & R6\\  \midrule
0 & temp sea pressure rain weather & .008 & .009 & .012 & .012 & .011 & .011 \\
1 & job check nurse advisor ref & .002 & .004 & .006 & .003 & .006 & .006 \\
2 & data storage file holiday song & .004 & .007 & .011 & .007 & .006 & .007 \\
3 & morning good kevin steve vacancy & .019 & .013 & .015 & .018 & .014 & .014 \\
4 & live saturday stream masquerade laugh & .011 & .010 & .009 & .008 & .007 & .004 \\
5 & jump long pit radio runway & .002 & .005 & .004 & .005 & .002 & .002 \\
6 & pitch synthetic turf artificial sport & .006 & .008 & .006 & .008 & .009 & .007 \\
7 & market sign september risk easy & .000 & .000 & .000 & .000 & .000 & .000 \\
8 & consultant resin sport pitch flooring & .009 & .012 & .013 & .008 & .007 & .008 \\
9 & aquarius sensational today seventy happen & .017 & .021 & .019 & .016 & .019 & .022 \\
10 & playground marking key stage game & .012 & .013 & .009 & .017 & .011 & .013 \\
11 & gallery collection art contemporary home & .006 & .002 & .002 & .003 & .002 & .002 \\
12 & cancer mouth breast research today & .005 & .008 & .010 & .007 & .008 & .008 \\
13 & safety train air cylinder pneumatic & .019 & .021 & .019 & .020 & .020 & .017 \\
14 & trade wale choose big car & .006 & .005 & .004 & .005 & .004 & .006 \\
15 & beautiful cute look amaze adorable & .066 & .057 & .063 & .077 & .072 & .075 \\
16 & manager director yoga technical executive & .008 & .011 & .007 & .006 & .008 & .011 \\
17 & course training certificate lunch click & .005 & .007 & .004 & .004 & .004 & .005 \\
18 & news north northern west warrior & .040 & .042 & .045 & .045 & .050 & .043 \\
19 & mobility product salary showroom look & .002 & .002 & .002 & .002 & .003 & .003 \\
20 & china chinese outbreak congo measles & .022 & .021 & .022 & .023 & .023 & .021 \\
21 & ref level surfacing sale representative & .009 & .010 & .013 & .009 & .009 & .007 \\
22 & interview job excellent benefit tip & .001 & .002 & .003 & .003 & .002 & .002 \\
23 & truck law year new minute & .015 & .017 & .013 & .011 & .008 & .008 \\
24 & privacy place data speaker security & .011 & .009 & .010 & .011 & .012 & .011 \\
25 & rule entry voucher submit year & .007 & .008 & .007 & .005 & .007 & .006 \\
26 & late daily predator bullet pip & .001 & .002 & .002 & .003 & .003 & .003 \\
27 & twitter mention reach week like & .019 & .018 & .020 & .021 & .022 & .018 \\
28 & birthday happy hope soon wish & .003 & .003 & .004 & .005 & .005 & .006 \\
29 & cheer agree true baby mate & .004 & .005 & .004 & .004 & .004 & .004 \\
30 & hockey court tennis final surface & .003 & .002 & .002 & .002 & .004 & .002 \\
31 & free instant horse audit tip & .001 & .000 & .000 & .000 & .001 & .000 \\
32 & branch rate store available buy & .000 & .000 & .000 & .000 & .000 & .000 \\
33 & support help child people sport & .011 & .017 & .013 & .010 & .014 & .014 \\
34 & tropical storm thunderstorm weather rain & .010 & .009 & .009 & .010 & .012 & .009 \\
35 & sleep bed night nap asleep & .003 & .002 & .002 & .004 & .002 & .002 \\
36 & property bedroom tax station family & .018 & .011 & .005 & .004 & .004 & .005 \\
37 & ship cruise new marine boat & .019 & .028 & .029 & .033 & .037 & .033 \\
38 & new star review unit charge & .003 & .004 & .003 & .003 & .003 & .004 \\
39 & wine competition enter medal sommelier & .004 & .004 & .003 & .006 & .004 & .005 \\
40 & cost value low decision help & .007 & .007 & .008 & .010 & .009 & .007 \\
41 & coffee tea cup lunch grandma & .003 & .004 & .003 & .004 & .004 & .005 \\
42 & christmas merry gift year festive & .047 & .041 & .052 & .054 & .042 & .051 \\
43 & click link workshop business poetry & .040 & .040 & .043 & .040 & .046 & .042 \\
44 & night tonight bar drink beer & .006 & .005 & .006 & .006 & .007 & .006 \\
45 & number guide model mary information & .033 & .022 & .015 & .016 & .010 & .012 \\
46 & cat kitten bruce love like & .049 & .037 & .053 & .052 & .054 & .057 \\
47 & food farm production course eat & .002 & .005 & .001 & .001 & .002 & .001 \\
48 & attack data breach security user & .024 & .039 & .043 & .041 & .038 & .035 \\
49 & garden summer plant grow flower & .005 & .004 & .005 & .007 & .008 & .008 \\
    \end{tabular}
   \label{tab:topics-random2-1}
\end{table}

\begin{table}[!h]
\centering
\scriptsize
\center
\setlength{\tabcolsep}{0.5em}
\renewcommand{\arraystretch}{1.2}
    \begin{tabular}{llcccccc}
50 & dog puppy pet guide animal & .007 & .006 & .003 & .004 & .003 & .005 \\
51 & pizza chicken cheese meat sausage & .004 & .005 & .005 & .004 & .005 & .004 \\
52 & miss today gemini watch courtesy & .007 & .004 & .007 & .008 & .006 & .007 \\
53 & health mental cigarette tobacco cricket & .001 & .001 & .002 & .002 & .001 & .001 \\
54 & brain injury scientist researcher science & .000 & .002 & .001 & .000 & .001 & .001 \\
55 & climate green change carbon environmental & .022 & .009 & .009 & .005 & .007 & .006 \\
56 & fisherman fish beanie fishery marine & .005 & .001 & .002 & .001 & .002 & .001 \\
57 & photo learn range support publish & .015 & .018 & .018 & .016 & .019 & .017 \\
58 & follow automatically unfollowed check person & .001 & .003 & .002 & .002 & .002 & .002 \\
59 & movie game best funny hot & .001 & .000 & .001 & .001 & .001 & .000 \\
60 & happy car customer new trade & .004 & .004 & .004 & .006 & .005 & .005 \\
61 & shower black halloween dance look & .010 & .011 & .010 & .010 & .012 & .013 \\
62 & ray order edition release win & .026 & .029 & .028 & .030 & .030 & .035 \\
63 & tomorrow evening close support message & .006 & .005 & .005 & .007 & .004 & .005 \\
64 & ticket tour sale wait announce & .002 & .006 & .008 & .007 & .006 & .006 \\
65 & music album new single pic & .011 & .011 & .010 & .011 & .012 & .013 \\
66 & monday wednesday tuesday flight fly & .021 & .018 & .019 & .022 & .019 & .023 \\
67 & friday library thursday fact hub & .005 & .008 & .009 & .007 & .008 & .007 \\
68 & pisces scorpio virgo aries stop & .004 & .005 & .004 & .004 & .005 & .006 \\
69 & rugby story world news cup & .012 & .008 & .007 & .006 & .007 & .007 \\
70 & red player play win game & .003 & .003 & .004 & .002 & .002 & .002 \\
71 & football league weekend win round & .010 & .011 & .012 & .012 & .013 & .014 \\
72 & model age commercial shoot female & .014 & .013 & .008 & .005 & .004 & .004 \\
73 & trump donald president like america & .002 & .005 & .006 & .006 & .008 & .008 \\
74 & school bullying start change cover & .005 & .010 & .009 & .007 & .010 & .009 \\
75 & student need require math support & .002 & .003 & .002 & .002 & .002 & .003 \\
76 & congratulation award woman queen category & .006 & .007 & .007 & .008 & .007 & .006 \\
77 & sorry order address number hear & .005 & .002 & .002 & .002 & .001 & .001 \\
78 & update android store form creator & .005 & .005 & .005 & .005 & .008 & .006 \\
79 & day valentine today good happy & .019 & .014 & .013 & .012 & .013 & .010 \\
80 & cement airport retail duty travel & .002 & .002 & .002 & .002 & .003 & .002 \\
81 & month year week contract sunday & .006 & .012 & .009 & .008 & .007 & .005 \\
82 & business parent feature social medium & .003 & .004 & .003 & .002 & .002 & .002 \\
83 & bank payment launch platform banking & .009 & .008 & .006 & .006 & .007 & .006 \\
84 & week hour month image shot & .017 & .017 & .015 & .020 & .016 & .018 \\
85 & leadership development network skill leader & .003 & .005 & .005 & .005 & .005 & .003 \\
86 & hug paw love send david & .003 & .005 & .003 & .003 & .004 & .003 \\
87 & police man old jail arrest & .012 & .012 & .010 & .010 & .010 & .010 \\
88 & road car driver vehicle cyclist & .012 & .010 & .008 & .008 & .008 & .008 \\
89 & car race drive raceway driver & .017 & .014 & .013 & .009 & .011 & .006 \\
90 & vote boris deal labour party & .001 & .002 & .002 & .002 & .003 & .004 \\
91 & tip time management try start & .006 & .010 & .011 & .007 & .009 & .016 \\
92 & address send password congratulation number & .007 & .010 & .009 & .007 & .007 & .008 \\
93 & bad hate sad sorry wrong & .004 & .005 & .005 & .005 & .006 & .006 \\
94 & subscription address sorry hear look & .001 & .001 & .002 & .001 & .001 & .002 \\
95 & echo team sorry touch order & .003 & .002 & .002 & .002 & .004 & .004 \\
96 & dont think know game like & .011 & .011 & .005 & .006 & .004 & .005 \\
97 & love sun island change book & .002 & .003 & .003 & .004 & .003 & .005 \\
98 & big follower fan week share & .010 & .005 & .006 & .006 & .007 & .007 \\
99 & yes yeah xmas amen everyday & .007 & .005 & .009 & .007 & .006 & .004 \\
        \bottomrule
    \end{tabular}
   \label{tab:topics-random2-2}
\end{table}

\end{document}